\newif\ifsingle

\newif\ifFullversion
\Fullversiontrue 

\ifsingle
\documentclass[11pt,draftclsnofoot, onecolumn]{IEEEtran}		
\else		
\documentclass[9pt,final, twocolumn]{IEEEtran}
\fi

\usepackage{bbm}
\usepackage{times}
\usepackage{amsmath,dsfont}
\usepackage{amssymb,amsthm}
\usepackage{epsfig,verbatim}
\usepackage{subcaption}
\usepackage{setspace}
\usepackage{color}
\usepackage{cite}
\usepackage{epstopdf}
\usepackage{graphics}
\usepackage{accents}
\usepackage{acronym}
\usepackage[bookmarks,colorlinks]{hyperref}
\usepackage{booktabs}
\usepackage{mathtools}
\usepackage{enumitem}
\usepackage{multirow}
\usepackage{booktabs}
\usepackage{hyperref}
\usepackage{epstopdf} 

\usepackage{float}
\usepackage[ruled,linesnumbered]{algorithm2e}  
\SetKwInput{KwData}{\textbf{Init}} 
\usepackage{algpseudocode} 

\newcommand{\myVec}[1]{{\boldsymbol{#1}}}

\definecolor{NewColor}{rgb}{0,0,0} 

\ifsingle

\else

\fi 

\acrodef{adc}[ADC]{analog-to-digital convertor}
\acrodef{cs}[CS]{compressed sensing}
\acrodef{dtft}[DTFT]{discrete-time Fourier transform}
\acrodef{dnn}[NN]{neural network} 
\acrodef{csi}[CSI]{channel state information}
\acrodef{map}[MAP]{maximum a-posteriori probability}
\acrodef{snr}[SNR]{signal-to-noise ratio}
\acrodef{bs}[BS]{base station} 
\acrodef{iot}[IOT]{Interent of Things}
\acrodef{mimo}[MIMO]{multiple-input multiple-output}
\acrodef{mse}[MSE]{mean-squared error}
\acrodef{pdf}[PDF]{probability density function}
\acrodef{rv}[RV]{random variable}
\acrodef{ml}[ML]{machine learning}
\acrodef{mf}[MF]{matched filter}
\acrodef{fec}[FEC]{forward error correction}
\acrodef{rs}[RS]{Reed-Solomon}
\acrodef{lti}[LTI]{linear time-invariant}
\acrodef{wss}[WSS]{wide-sense stationary}
\acrodef{psd}[PSD]{power spectral density}
\acrodef{ser}[SER]{symbol error rate} 
\acrodef{ber}[BER]{bit error rate} 
\acrodef{sgd}[SGD]{stochastic gradient descent} 
\acrodef{isi}[ISI]{intersymbol interference}  
\acrodef{awgn}[AWGN]{additive white Gaussian noise} 
\acrodef{ut}[UT]{user terminal} 
\acrodef{mmw}[mmWave]{millimeter wave}
\acrodef{noma}[NOMA]{non-orthognal multiple access}
\acrodef{mac}[MAC]{mulitple access channel}
\acrodef{fl}[FL]{federated learning}
\acrodef{ct}[CT]{continuous-time}

\IEEEoverridecommandlockouts

\title{LIDAR and Position-Aided mmWave Beam Selection with Non-local CNNs and Curriculum Training}
\author{ Matteo Zecchin$^{*}$, Mahdi Boloursaz Mashhadi$^{*}$, Mikolaj Jankowski$^{*}$, Deniz G\"{u}nd\"{u}z, Marios Kountouris, David Gesbert \thanks{$^*$ Equal contribution.\\
M. Zecchin, M. Kountouris and D. Gesbert are with the Communication Systems Department, EURECOM, Sophia-Antipolis, France. M. Boloursaz Mashhadi is with the 5GIC \& 6GIC, Institute for Communication Systems (ICS), University of Surrey, UK, and M. Jankowski and D. G\"{u}nd\"{u}z are with the Dept. of Electrical and Electronic Eng., Imperial College London, UK.\\
The work of M. Zecchin is founded by the Marie Sklodowska Curie action WINDMILL (grant No. 813999). Mikolaj Jankowski and D. Gunduz received funding from the European Research Council through project BEACON (grant  No. 677854). D. Gesbert's and M. Kountouris's contribution are partially supported from a Huawei France-funded Chair towards Future Wireless Networks.}}

\vspace{-0.75cm}

\begin{document}
	
	\maketitle
	\pagestyle{empty}
	\thispagestyle{empty}
	
\begin{abstract}
Efficient millimeter wave (mmWave) beam selection  in vehicle-to-infrastructure (V2I) communication is a crucial yet challenging task due to the narrow mmWave beamwidth and high user mobility. To reduce the search overhead of iterative beam discovery procedures, contextual information from  light detection and ranging (LIDAR) sensors mounted on vehicles has been leveraged by data-driven methods to produce useful side information. In this paper, we propose a lightweight neural network (NN) architecture along with the corresponding LIDAR preprocessing, which significantly outperforms previous works.  
Our solution comprises multiple novelties that improve both the convergence speed and the final accuracy of the model. In particular, we define a novel loss function inspired by the knowledge distillation idea, introduce a curriculum training approach exploiting line-of-sight (LOS)/non-line-of-sight (NLOS) information, and we propose a non-local attention module to improve the performance for the more challenging NLOS cases. Simulation results on benchmark datasets show that, utilizing solely LIDAR data and the receiver position, our NN-based beam selection scheme can achieve 79.9\% throughput of an exhaustive beam sweeping approach without any beam search overhead and 95\% by searching among as few as 6 beams. {\color{black}In a typical mmWave V2I scenario, our proposed method considerably reduces the beam search time required to achieve a desired throughput, in comparison with the inverse fingerprinting and hierarchical beam selection schemes.}



\end{abstract}

\section{Introduction}\label{sec:intro}

Millimeter wave (mmWave) communication constitutes a fundamental technology in 5G and future networks, which allows to overcome communication bottlenecks of the over-exploited sub-6GHz bands. To overcome the severe propagation impairments of the above-10GHz spectrum, such as high path attenuation and penetration losses, mmWave communication systems employ massive number of antennas at the base station (BS) to form highly directional beams and attain a large beamforming gain.  Because of the narrow mmWave beamwidth, extremely precise alignment and tracking procedures are necessary in order to establish a reliable and high throughput communication link. The optimal communication beam can be easily determined with full channel knowledge; however, in the large antenna regime, obtaining an estimate of the high dimensional channel matrix is costly; and hence, beam selection for efficient communication requires iterative search procedures. In vehicular-to-infrastructure (V2I) communications, for which mmWave communication is envisioned to be a key technology \cite{choi2016millimeter}, beam selection and tracking are particularly challenging due to the high mobility of the receivers, which leads to reduced beam coherence time \cite{va2016impact}. In this scenario, conventional beam selection techniques, such as beam sweeping or multi-level beam selection \cite{wei2018initial,hur2011multilevel}  impose a significant overhead. Therefore, more efficient beam selection techniques that can reduce the cost of iterative search procedure by exploiting contextual information are of great interest.

It has been shown that contextual information from sensors mounted on the vehicles and the infrastructure can be leveraged to reduce the beam selection overhead. For instance, the position information provided by vehicle global positioning system (GPS) can be used to apply an inverse fingerprint approach and query the most prominent mmWave beams \cite{1}. Inertial sensors placed on vehicle's antenna arrays enable efficient antenna element configuration by tracking the orientation of the vehicle \cite{brambilla2019inertial}. Furthermore, positional and motion information can be jointly processed to further reduce the alignment overhead \cite{8277251}. From the infrastructure side, a radar located at the BS can help estimate the direction of arrival, which would aid the beam search \cite{7}. Spatial information obtained from out of band measurements was exploited in \cite{5, 6, Sub6G1, Sub6G2} where \cite{Sub6G1, Sub6G2} used sub-6GHz channel measurements to train neural network (NN)s for mmWave beamforming. Vision-aided approaches were proposed in \cite{vision, alrabeiah2020millimeter, Alkhateeb}. Base stations equipped with cameras were proposed to employ computer vision and deep learning techniques to predict mmWave blockage and beam strength in  \cite{alrabeiah2020millimeter}. The authors in \cite{Alkhateeb} built a panoramic point cloud from images taken within the cellular coverage area. This point cloud gives a view of the scattering environment, which is then input to a neural network (NN) to predict the optimal beams.

Thanks to recent surge of autonomous driving technologies, high dimensional sensor information is nowadays commonly available also at the vehicle side. For instance, light detection and ranging (LIDAR) is commonly used for autonomous navigation. LIDAR uses a laser to produce a depth map of the environment and surrounding obstacles using delay measurements of the back-scattered signal. Because of the data dimensionality and the lack of analytical models that would relate LIDAR depth map to mmWave beams quality, data-driven methods have been considered to effectively process LIDAR signals as side information for beam search. In \cite{8, 9}, a NN architecture was trained over simultaneous LIDAR and ray-tracing channel datasets with a top-$k$ classification metric to identify $k$ beam directions that most probably include the beam resulting in the largest channel gain. In order to reduce the computational cost and NN model size, a simplified classifier architecture that can be trained in a distributed fashion using federated learning was proposed in \cite{IPC}. 




This paper builds on the unpublished work of the authors that recently won the ``AI/ML in 5G" competition ranking \textit{first} in the ``ML for mmWave beam selection" challenge organized by the International Telecommunications Union (ITU) \cite{challenge, ranking}. We propose a convolutional neural network (CNN) architecture along with the corresponding LIDAR preprocessing technique for data-driven mmWave beam selection. The proposed model is trained to exploit LIDAR and positional data in order to identify the best beam directions and reduce the beam search overhead in V2I communication. The specific contributions of this paper in comparison with previous works \cite{8, 9, IPC} can be summarized as follows: 
\begin{itemize}[leftmargin=*,leftmargin=\dimexpr\labelwidth + 2\labelsep\relax]
    \item Inspired by the knowledge distillation (KD) techniques \cite{hinton2015distilling}, we propose a novel loss function, which not only maximizes the prediction accuracy of the best beam index, but also its corresponding  power gain. The proposed loss function improves the beam prediction accuracy specifically for smaller $k$ values achieving considerably higher throughput with significantly reduced beam search overhead.

    \item We utilize a non-local attention scheme, which improves the beam classification accuracy, specifically for the non-of-sight (NLOS) case. Convolutional classifiers used in previous works \cite{8, 9 , IPC} learn local features from the LIDAR input and exploit them for beam classification. We observe that the NN utilizing our proposed non-local attention module considerably benefits from a non-local perception of the LIDAR input, specifically in NLOS scenarios where the mmWaves may be reflected from scatterers located far away.

    \item We propose a curriculum training strategy, which improves both the convergence speed and the final beam prediction accuracy. We observe that for the samples with no dominant LOS component, the strongest propagation path becomes significantly less predictable as it depends on the location of scatterers and reflectors, which is mainly determined by the traffic conditions. With NLOS samples being more challenging, the proposed curriculum learning strategy starts training with the LOS samples first, and gradually exposes the classifier to more complex NLOS samples. This training strategy achieves faster convergence and improved beam classification accuracy.
    
    \color{black}
    \item We also present a protocol to integrate the proposed NN-based beam alignment scheme into the existing fifth generation (5G) new radio (NR) standard. We provide comparisons with the hierarchical beam selection \cite{noh2017multi} and inverse fingerprinting \cite{va2017inverse} protocols in terms of the time required for beam search. The simulation results show that, in spite of a short initial processing delay, the proposed solution outperforms alternative beam alignment schemes in terms of the achieved throughput ratio. In a typical mmWave V2I scenario, our proposed method considerably reduces the beam search time required to achieve a desired throughput, in comparison with the inverse fingerprinting and hierarchical beam selection schemes.
    \color{black}
    
\end{itemize}

Thanks to the above ideas, our NN classifier significantly outperforms previous works \cite{8, 9, IPC}. Utilizing the benchmark Raymobtime dataset \cite{raymobtime, raymobtime1}, our solution achieves top-1, top-5 and top-10 beam selection accuracies of 59.5\%, 87.0\%, and 92.2\%, respectively. In a mmWave communication system with 256 possible beam pairs, our LIDAR-based approach achieves $95\%$ of the available throughput, only by searching among the 6 most probable beams suggested by the NN classifier, greatly reducing the beam search space and the corresponding beam selection overhead. Our classifier harnesses, on average, 79.9\% of the available throughput without any beam search at all, {\color{red}just by utilizing the LIDAR and position information.} Finally, we show that the proposed NN classifier can be further simplified utilizing effective NN pruning techniques without significant loss of its performance while reducing the computational and storage costs for practical deployment.

The content of the paper is organized as follows: Sec.~\ref{sec:Model} introduces the system model. Sec.~\ref{sec:Approach} illustrates our proposed NN model for mmWave beam selection utilizing LIDAR data. Simulation results are reported in Sec.~\ref{sec:results}. {\color{red}Sec.~\ref{protocol} presents a protocol for NN-based beam alignment to be integrated into the existing 5G NR standard.} Finally, Sec.~\ref{sec:Conclusions} concludes the paper. The simulation code is publicly available at: \href{https://github.com/MatteoEURECOM/LIDAR-mmWave-Beam-Selection}{https://github.com/MatteoEURECOM/LIDAR-mmWave-Beam-Selection}. 

\section{System Model}\label{sec:Model}
We consider a downlink orthogonal frequency-division multiplexing (OFDM) mmWave system using analog beamforming, where the BS located on the street curb serves a vehicle in its coverage area utilizing $N_c$ subcarriers. Both the transmitter and the receiver ends are equipped with antenna arrays with a single radio frequency (RF) chain and fixed complex beam codebooks, which we denote by $\mathcal{C}_t = \{\mathbf{f}_i\}_{i=1}^{C_t}$ and
$\mathcal{C}_r = \{\mathbf{w}_j\}_{j=1}^{C_r}$, respectively. The downlink channel matrix from the BS to the vehicle over the $n$'th subcarrier is denoted by $\mathbf{H}_n$. 

\begin{figure*}
    \centering
    \includegraphics[scale=.7]{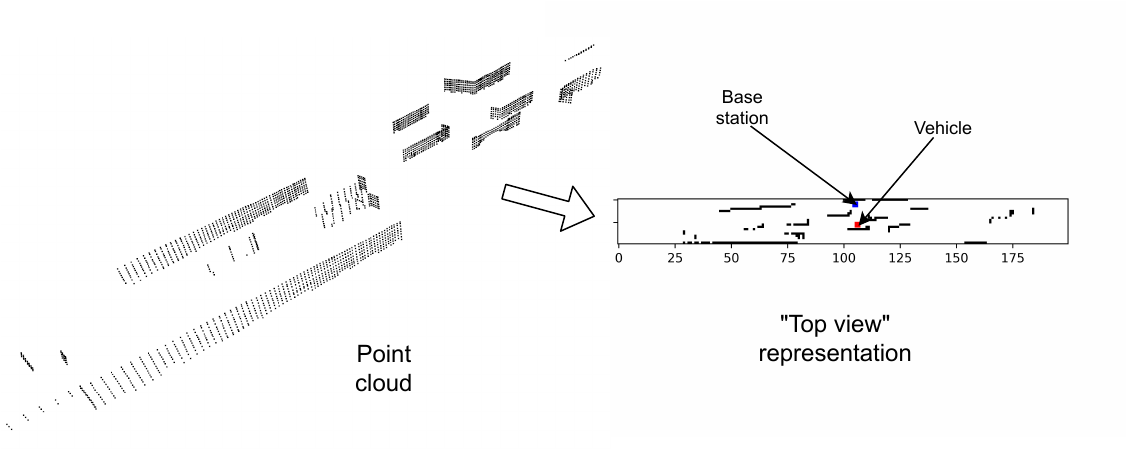}
    \caption{Preprocessing of the LIDAR point cloud.}
    \label{fig:pcl22d}
\end{figure*}

\begin{figure*}
    \centering
    \includegraphics[width=\linewidth]{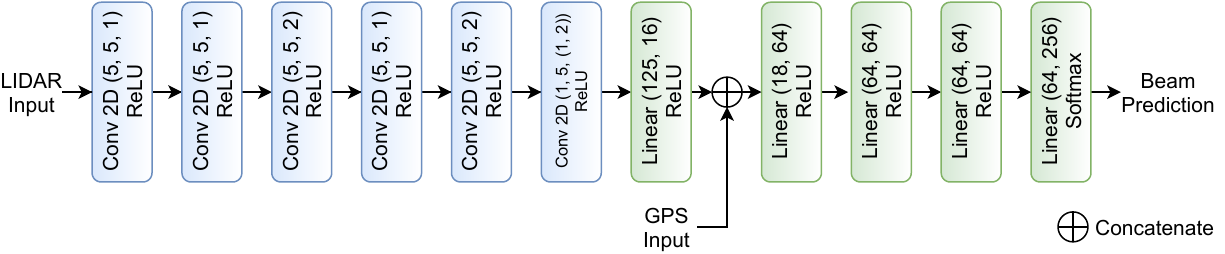}
    \caption{The proposed CNN model architecture.}
    \label{fig:architecture}
\end{figure*}

For each precoder and combiner vector pair $(i, j) \in \mathcal{C}_t \times \mathcal{C}_r$, the resulting channel gain at subcarrier $n$ is determined by $\mathbf{w}_j^H \mathbf{H}_n\mathbf{f}_i$, where $(\cdot)^H$ denotes the conjugate transpose. {\color{black}Without loss of generality, we assume a unit transmit power over each subcarrier and a unit noise variance. The receive power summed over all subcarriers for the transmitter-receiver codebook pair $(i,j)$ is 

\begin{align}\label{powergain}
    \mathbf{G}_{i,j} =\sum_{n=1}^{N_c} |\mathbf{w}_j^H \mathbf{H}_n\mathbf{f}_i|^2,
\end{align}
and the optimal pair of precoding and combining vectors that maximizes the receive signal-to-noise ratio (SNR) is given by 

\begin{equation}
    (i^*,j^*)=\underset{(i,j)}{\operatorname{argmax}}\  \mathbf{G}_{i,j}.
\end{equation}}

Without side information, the transmitter and receiver need to perform an exhaustive search through all $C_t\times C_r$ beam pairs in order to identify $(i^*,j^*)$. Our goal is to infer a small subset of $k$ beam pairs $\mathcal{S}_k \subset \mathcal{C}_t \times \mathcal{C}_r$  exploiting the available position and LIDAR data, such that
$(i^*,j^*) \in \mathcal{S}_k$. This results in a reduction of $\frac{k}{C_t \times C_r}$ in the search space of the beam selection procedure. Two metrics to gauge the quality of $\mathcal{S}_k$ as a function of its size $k$, are the top-$k$ accuracy and top-$k$ throughput ratio. The top-$k$ accuracy is formally defined as 
\begin{equation}
    \label{eqn:topk}
    A(k)=\mathbb{E}\left[\mathbbm{1}\{(i^*, j^*)\in \mathcal{S}_k\}\right],
\end{equation}
{\color{black}where $\mathbbm{1}\{(i^*, j^*)\in \mathcal{S}_k\}$ is the indicator function associated to the event that the best beam index is in the top-$k$ classifier output.}
{\color{black}With the unit noise variance and transmit power assumptions, the top-$k$ throughput ratio defined as 
\begin{equation}
    \label{eqn:throughput}
    T(k)=\frac{\mathbb{E}\left[\max_{(i,j)\in\mathcal{S}_k }\log_2(1+\mathbf{G}_{i,j})\right]}{\mathbb{E}\left[\log_2(1+\mathbf{G}_{i^*, j^*})\right]}
\end{equation}
is used as a complementary performance metric, where all expectations are with respect to the inherent randomness introduced by the vehicles' positions, channel realizations, and LIDAR measurements.} Note that the top-$k$ throughput ratio is a very informative metric for the problem at hand. In fact, the numerator represents the throughput that can be achieved (at a zero dB transmit SNR) by searching only among the top-$k$ beams suggested by the NN; while the denominator is a normalizing factor representing the maximum throughput achievable by an exhaustive beam sweeping approach.

 In the next section, we propose a novel NN architecture that  jointly processes location information along with LIDAR data in order to solve the top-$k$ classification task; and therefore, to find the most promising mmWave beams to establish a reliable communication link with reduced beam search overhead.

\section{CNN-based Beam Selection Utilizing LIDAR Data}\label{sec:Approach}
We propose a novel CNN-based beam selection scheme, where connected vehicles utilize measurements from their LIDAR sensors along with their location data to reduce the beam search overhead required to establish a mmWave link with a nearby  BS. 

\subsection{LIDAR Prepossessing}
\label{subsec:lidar_preprocessing}

Raw LIDAR data is in the form of large point cloud measurements $\mathcal{P}=\{(x_i, y_i, z_i)\}_{i=1}^{|\mathcal{P}|}$, where each triplet $(x_i, y_i, z_i)$ represents coordinates of an obstacle point measured by the LIDAR sensor. To avoid excessive computations on large point clouds, we preprocess the raw LIDAR data to get a simplified representation of the coverage area of the BS, which is then input to the classifier CNN. We assume that each vehicle knows the location of the BS and its coverage area, and divides the coverage area into a 2D grid of equal squares. We then produce a top-view representation of the coverage area setting grid entries to 1 whenever at least one point in $\mathcal{P}$ lies within that grid square, and to 0 otherwise. We also embed the location of the vehicle and the BS into this representation by setting the grid value of the square accommodating the BS and the vehicle to $-1$ and $-2$, respectively. Fig. \ref{fig:pcl22d} shows one such preprocessing step. We note that discarding the $z$-axis causes certain information loss; however, we found this loss not to affect the accuracy of the beam classification task considerably. On the other hand, it allows us to reduce the complexity of our NN model significantly.

\subsection{NN Architecture}
\label{subsec:Architecture}

Fig. \ref{fig:architecture} shows our proposed NN architecture for LIDAR-aided mmWave beam selection, which is composed of 6 convolutional layers and 5 linear layers. Each layer is followed by the rectified linear unit (ReLU) activation, except for the last layer, which is followed by softmax activation to output beam predictions. Our architecture consists of separate branches to process LIDAR and GPS inputs. The LIDAR branch comprises 6 convolutional layers followed by a linear layer to extract features from the preprocessed LIDAR input as in
Sec. \ref{subsec:lidar_preprocessing}. The output features from the LIDAR branch are then concatenated with $(x,y)$ location coordinates of the vehicle from the GPS input. We discard the BS coordinates, as these are fixed on the whole dataset and do not need to participate in the training process. 

The concatenated feature vector is then input to 4 linear layers. The first three linear layers include 64 neurons and the last one outputs a 256-element vector corresponding to the number of possible beam pairs, i.e., $|C_t|\cdot|C_r|=256$. Although these linear layers increase the complexity of our NN architecture, in Sec. \ref{subsection:pruning} we use pruning techniques  to effectively reduce the memory and computation requirements of the proposed NN. 

\subsection{Loss Function}
\label{subsec:CostFunction}
\color{black}
To define the loss function, we denote by $\mathbf{y} \in \mathbbm{R_+}^{(|\mathcal{C}_t| \cdot |\mathcal{C}_r|)}$ the vectorized version of the mmWave power gain matrix $\mathbf{G}$ obtained by the following bijective map
\begin{equation}
\mathbf{y}_{(i-1)\cdot|\mathcal{C}_r|+j}=\mathbf{G}_{i,j}.
\end{equation}
For each  $\mathbf{y}$ vector, we also denote by $\bar{\mathbf{y}}$ the vector $\mathbf{y}$ normalized to unit  Euclidean norm and by $\mathbf{y}^*$ the unitary vector that is non-zero on the component corresponding to the largest entry of $\mathbf{y}$ (ties  broken arbitrarily). Then, for $\beta\in [0,1]$, we train our model by minimizing the following loss function

\begin{equation}
\ell(\mathbf{y},\hat{\mathbf{y}})=(1-\beta)\mathcal{H}(\mathbf{y}^*,\hat{\mathbf{y}})+\beta\mathcal{H}(\bar{\mathbf{y}},\hat{\mathbf{y}})
\label{eqn:KDloss}
\end{equation}
where $\hat{\mathbf{y}}$ is the model prediction and $\mathcal{H}(\cdot)$ denotes the empirical cross entropy that, for two non-negative unit norm vectors $p$ and $q$ in $\mathbb{R}^d$ is defined as
\begin{equation}
\mathcal{H}(p,q)=-\sum^d_{i=1}p_i\log(q_i).
\label{eqn:CE}
\end{equation} The first term in (\ref{eqn:KDloss}) is a standard multi-class cross entropy loss that enforces the NN to predict the indices of the beam associated to the strongest mmWave beam. The second term, instead, drives the NN to match the output of each neuron in the last layer to the normalized power gain of the corresponding mmWave beam pair. This is achieved by treating $\mathbf{y}$ as a vector of ``soft labels" and minimizing the corresponding empirical cross entropy loss.  This last term is specifically effective in top-$k$ classification for $k>1$, where it is advisable not to output exclusively the best beam pair, but also accurately predict $k$ competing candidate beam pairs. Finally, the value $\beta \in [0,1]$ provides a tradeoff between the  two terms. The idea of combining two training objectives as in Eq. (\ref{eqn:KDloss}) resembles the KD technique \cite{hinton2015distilling}. KD is a popular model compression technique that aims at instilling the knowledge of a large classifier network, termed as \textit{teacher}, into a lightweight \textit{student} classifier. This is achieved by augmenting the original labels of a dataset by the soft prediction of the teacher model. This additional training objective has been shown to improve the performance of the student model, in some cases even outperforming the teacher model, and to act as a regularization term \cite{hinton2015distilling,KD2}. In the context of mmWave beam selection, we show experimentally that by exploiting  the soft labels, we are able to obtain similar gains and improve the predictive capabilities of the trained model.


\subsection{Curriculum Training with LOS/NLOS Samples} \label{sec:Curriculum}
In the absence of LOS, the predictability of the strongest propagation paths greatly decreases as a consequence of the prominent dependency on the relative positions of scatterers and reflectors. However, the presence of neighbouring moving obstacles renders NLOS condition frequent in vehicular type of communication. The difficulty of the prediction task in NLOS scenarios is so high compared to the LOS case that, data-driven methods tend to be biased towards the easier LOS samples to the detriment of the NLOS performance. In order to address this performance imbalance, we propose to adapt the sample distribution during training time so that the challenging instances are less likely during the initial phases of learning. This procedure is motivated by curriculum learning \cite{bengio2009curriculum, Curriculum2}, which suggests to expose the training process to easier instances at the initial phases and to gradually increase the difficulty of the tasks. To apply this strategy it is necessary to first define \textit{scoring} and a \textit{pacing} functions: the former assigns a level of difficulty to each sample while the latter determines at which rate the transition should be made from easier samples to harder ones during the learning process.

For the task at hand, a natural measure of difficulty is the absence of LOS and an effective way to modulate the difficulty of the learning task consists in changing the probability of NLOS samples during the training epochs. Hence, denoting by $P$ the feature distribution from which the original training dataset $\mathcal{D}$ is generated, we exploit a biased sampling scheme to generate skewed dataset $\mathcal{D}_\lambda$, whose hardness is proportional to the rejection coefficient $\lambda\in [0,1]$.  In particular, the set of instances $\mathcal{D}_\lambda$ is created from $\mathcal{D}$ by independently removing each NLOS sample with probability $1-\lambda$. As a result, $\mathcal{D}_\lambda$ represents a sample drawn from the following distribution
\begin{align}
    P_{\lambda}\propto(1-q)P_{LOS}+q\lambda P_{NLOS}
\end{align}
where $P_{LOS}$ is the feature distribution conditioned on the presence of LOS,  $P_{NLOS}$ the distribution conditioned on its absence, and $q$ is the probability  of NLOS condition under the original distribution $P$.
The pacing function is represented by a sequence $\{\lambda_i\}$ that for each epoch $i$ determines the probability that a NLOS sample is accepted for training. As shown below, a properly chosen sequence $\{\lambda_i\}$ can improve both the convergence speed and the accuracy of the final solution compared to the unstructured and randomized sampling of training instances.

\subsection{Non-local Attention for Improved NLOS Performance}\label{sec:NLA}
\begin{figure}[t!]
\centering
\includegraphics[scale=0.75]{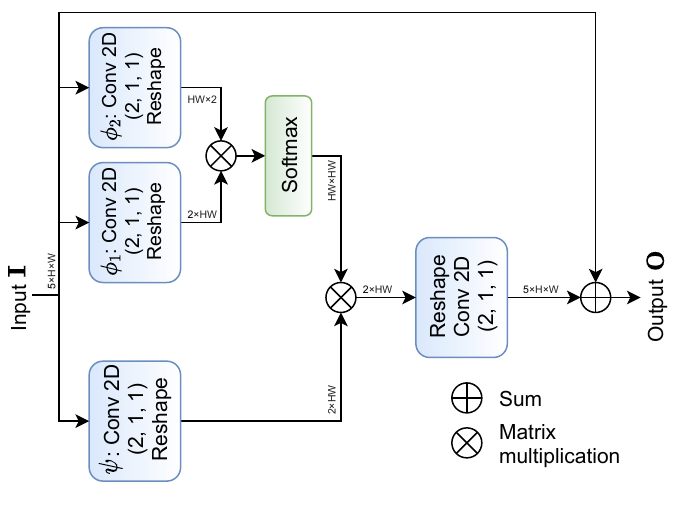}
\caption{Block diagram of the non-local attention module.}
\label{attentionBD}
\end{figure}

The convolutional layers in Fig. \ref{fig:architecture} learn and exploit local features from the LIDAR input and use them for beam classification. However, the classifier can benefit from non-local perception of the coverage area specifically in the NLOS cases. In order to extract non-local perception of the coverage area from the LIDAR input in an efficient manner, we use a non-local attention module introduced in \cite{attention1, attention2}. We later show through simulations that the non-local attention module further improves the performance. The general input-output relation for non-local attention is given by
\begin{align}\label{attention}
    \mathbf{O}_i=\mathbf{I}_i+\frac{1}{\eta(\mathbf{I})} \sum_{\forall j} \phi(\mathbf{I}_i, \mathbf{I}_j) \psi(\mathbf{I}_j),
\end{align}
in which $i$ is the space-time index of an output position whose response is to be computed, $j$ is the index that enumerates all possible positions, $\mathbf{I}$ is the input, and $\mathbf{O}$ is the output of the same size as $\mathbf{I}$. A pairwise function $\phi$ computes a scalar representing the relationship between $i$ and all $j$. The unary function $\psi$ computes a representation of the input signal at position $j$. The response is normalized by the factor $\eta(\mathbf{I})$. We tried various popular choices for $\eta(\cdot), \phi(\cdot)$  and $\psi(\cdot)$ functions (refer to \cite{attention2} for further details), and found that the best performing ones in our case are the embedded Gaussian function for $\phi(\mathbf{I}_i, \mathbf{I}_j)$, given by $\phi(\mathbf{I}_i, \mathbf{I}_j)=\exp{([\mathbf{W}_{\phi_1} \mathbf{I}_i]^T[\mathbf{W}_{\phi_2} \mathbf{I}_j])}$, $\psi(\mathbf{I}_j)=\mathbf{W}_{\psi} \mathbf{I}_j$, and $\eta(\mathbf{I})=\sum_{\forall j} \phi(\mathbf{I}_i, \mathbf{I}_j)$, as will be discussed later in Sec. \ref{sec:results}. Here, $\mathbf{W}$s are trainable weight matrices. Fig. \ref{attentionBD} provides the block diagram for our non-local attention module, where $1 \times 1$ convolutions implement $\mathbf{W}$ weight multiplications, and Softmax activation implements the exponential function. Refer to \cite{attention2} for more details on attention modules.

\subsection{Network Pruning}
\label{subsection:pruning}
To reduce the computational and memory footprint of the proposed model, we introduce an additional network pruning \cite{pruning, pruning2} step. Network pruning is a method for reducing the computational complexity and the size of NNs by finding redundant neurons and removing them, based on some saliency measure. In this work we employ a straightforward approach of removing neurons or filters with the lowest $L_1$-norm of the weights, which already leads to maintaining a satisfactory performance with low network complexity. We leave the exploration of more effective pruning methods for future work. Since the majority of the weights in our network are contained in the later linear layers the convolutional part of the network is already lightly parameterized. In this work we explore two particular strategies for network pruning: \textit{unstructured}and \textit{structured} pruning. Unstructured pruning removes the weights with the smallest magnitude, whereas structured pruning removes entire neurons with the lowest mean magnitude of weights. Unstructured pruning usually leads to better gains, as it is able to prune more parameters, while maintaining satisfactory performance. 

\section{Numerical Evaluations}\label{sec:results}
{\color{black}In this section, we provide a series of experiments highlighting the performance improvements and benefits that each of the proposed techniques and architecture designs can attain.} Subsequently, we evaluate the proposed solution with all of the above enhancements, compare it against the state of the art and showcase its superiority. For performance comparisons, we use the top-$k$ classification accuracy and the top-$k$ throughput ratio as defined in Sec. \ref{sec:Model}.

{\color{black}\subsection{Dataset} All experiments are carried out using the benchmark Raymobtime dataset \cite{raymobtime, raymobtime1}, which contains synthetic data for LIDAR and position-aided mmWave beam selection in a V2I communication scenario, where vehicles connect to a road side unit (RSU). The dataset is generated using a range of simulators including tools for transforming satellite images into realistic 3D urban scenes, mobility simulator for generating traffic data, and tools for generating ray-tracing mmWave channels. The resulting dataset contains the average received powers $\mathbf{G}_{i,j}$ calculated for each transmitter/receiver beam pair, together with the simultaneous 3D point cloud $\mathcal{P}$ collected by the LIDAR sensor. In this work, we utilize s008 and s009 subsets of the Raymobtime dataset, which are based on the data simulated for one of the streets in Rosslyn, Virginia. Each of the subsets considers a carrier frequency of 60GHz and contains 2086 and 2000 episodes (time-instances), respectively. The episodes were collected in the 30-second intervals of time. We train all models using s008, whereas we utilize s009 for testing.}


\begin{figure}
    \centering
    \includegraphics[width=\linewidth]{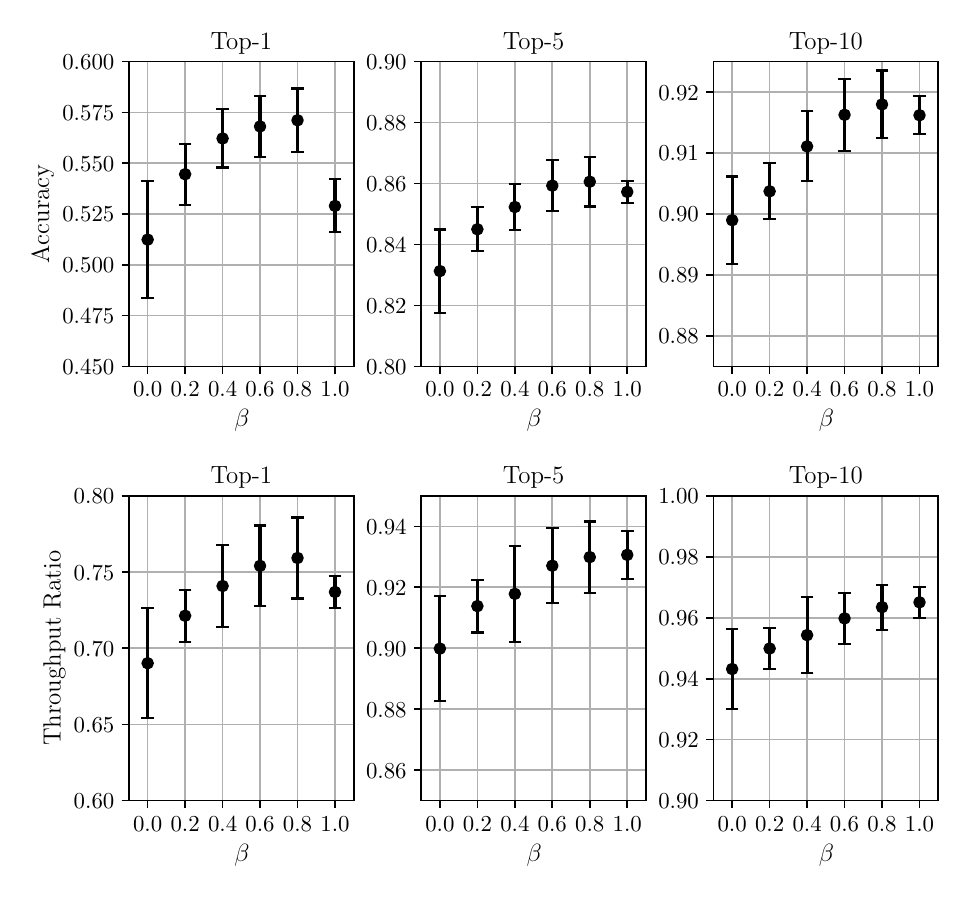}
    \caption{\color{black}{Average top-$k$ accuracy and througput ratio, with 95\% confidence intervals, for $k\in\{1,5,10\}$ averaged over 10 training runs for different $\beta$ values.}} 
    \label{fig:KDAccTH}
\end{figure}

\subsection{Choice of $\beta$}
We optimize the NN parameters by minimizing the proposed loss function (\ref{eqn:KDloss}). Note, however, that the proposed loss function is parameterized by $\beta\in [0,1]$ weighting the training signal coming from the hard and soft labels, respectively, represented by the one-hot-encoded best beam index $\mathbf{y}^*$ and the normalized channel gain vector $\bar{\mathbf{y}}$. Choosing a proper value for $\beta$ is necessary in order to strike a good balance between these two pulling forces. For small $\beta$ values, the objective function resembles the standard multi-class cross-entropy loss and the trained NN tends to act as a myopic classifier that tries to correctly predict the best beam index regardless of the performance of the other beams. On the other extreme, very large values of $\beta$ drive the output of the model to match the normalized channel gains associated with different  beams and to potentially trade the best beam prediction accuracy for this purpose. We evaluate the effect of $\beta$ by training the model proposed in Sec. \ref{subsec:Architecture} for $\beta\in \{0,0.2,0.4,0.6,0.8,1\}$ and we report the final top-$k$ accuracy and throughput ratio averaged over 10 runs for each $\beta$ value in Fig. \ref{fig:KDAccTH}. We observe that a value of $\beta=0.8$ yields a trained model that combines the best of the two above mentioned behaviours. In fact, for this choice of $\beta$, the predictor attains the highest top-1 accuracy, outperforming even the models trained for smaller $\beta$ values that prioritize this metric. At the same time, for larger $k$, its performance is indistinguishable from the network trained with $\beta=1$. The same conclusion holds for the top-$k$ throughput ratio metric. The model trained with the optimal $\beta$ value is able to provide 76\% top-$1$ throughput ratio, while the ratios for the two worst performing values are 69\% ($\beta=0$) and 73\% ($\beta=1$). 

{\color{black}Fig. \ref{fig:KDAccTH} also shows that incorporating in the loss function our proposed KD term not only improves the accuracy and throughput ratio, but with a proper choice of $\beta$ it also helps to reduce the variance. For example, without the second term (i.e., for $\beta=0$), the top-$1$ accuracy shows a relatively wider confidence interval. This means that, although we get a good average beam prediction accuracy, there exist settings in which our prediction accuracy is unsatisfactory. A tighter confidence interval ensures that such settings are infrequent. According to this figure, $\beta=0.8$ is an excellent choice that provides not only the best accuracy and throughput, but also a small variance.}

\subsection{Curriculum Training}

\begin{table}[t]
\centering
\caption{Performance of curriculum, anti-curriculum and standard training procedures.}
\begin{tabular}{|c|c|c|c|c|} 
\hline
Strategy & $A(1)$ & $A(5)$  & $T(1)$ & $T(5)$ \\ 
\hline
\hline
Curr. & \textbf{58.1\% }& \textbf{86.6\%}  & \textbf{77.9\%} &  \textbf{94.1\%} \\
\hline
Standard & 57.1\% & 86.0\%  & 75.9\%& 92.9\% \\
\hline
Anti-curr. &  53.8\% & 85.1\% & 71.9\% & 91.2\%  \\
\hline
\end{tabular}
\vspace{1em}
\label{tab:Curr}
\end{table}

\begin{figure*}[!htb]
\centering
\minipage{0.35\textwidth}
  \includegraphics[width=\linewidth]{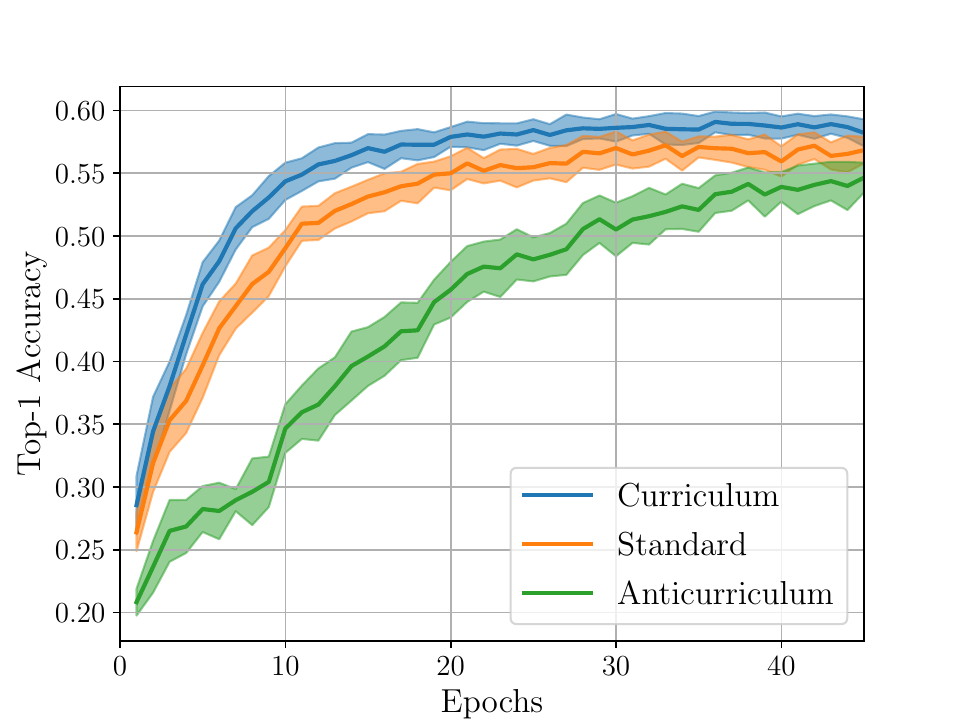}
  \subcaption{$k=1$}
\endminipage
\hspace{-7mm}
\minipage{0.35\textwidth}
  \includegraphics[width=\linewidth]{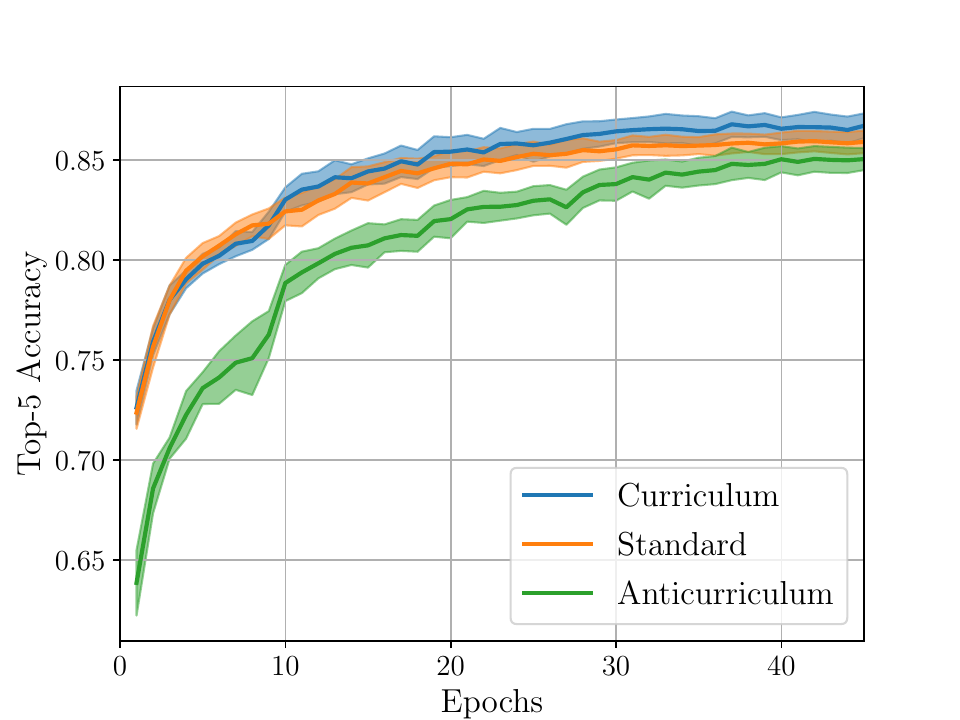}
  \subcaption{$k=5$}
\endminipage
\hspace{-7mm}
\minipage{0.35\textwidth}%
  \includegraphics[width=\linewidth]{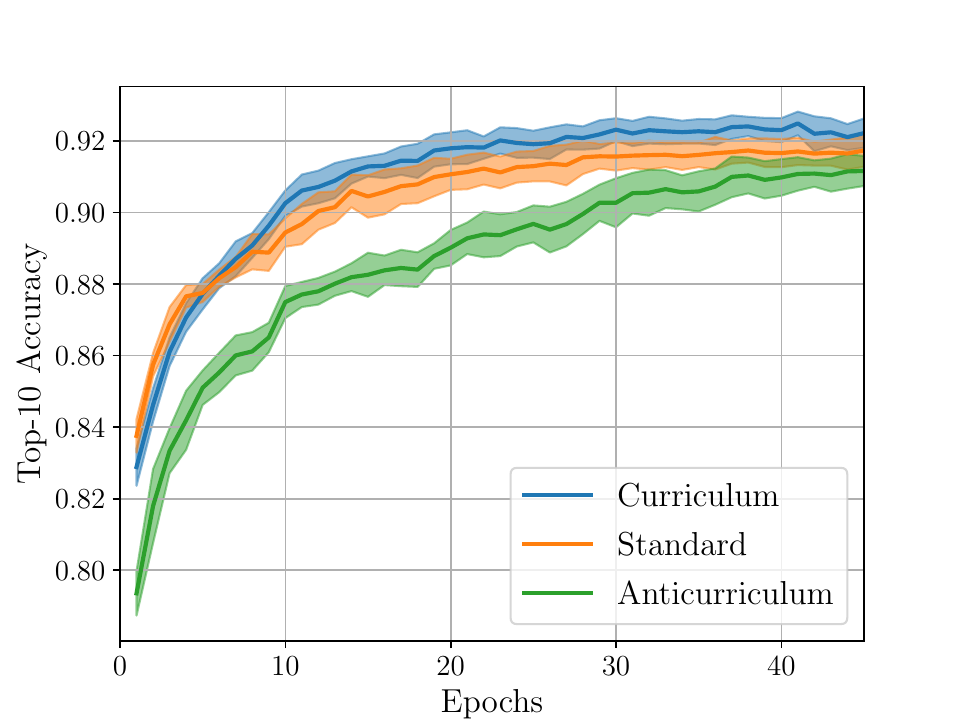}
  \subcaption{$k=10$}
\endminipage
\caption{Evolution of the top-$k$ accuracy over the training epochs for curriculum, standard and anti-curriculum training approaches. The curves are averaged over 10 training runs.}\label{fig:CurrLearningCurves}
\end{figure*}
 
While it is usual practice to train NNs using batches of data sampled uniformly at random from the training dataset, in the following we illustrate the benefits of biasing the sampling procedure in order to obtain a sequence of training samples with an increasing level of difficulty. As illustrated in Sec. \ref{sec:Curriculum}, this is achieved by employing a sample rejection strategy based on the presence or absence of LOS components.  Specifically, we train the proposed NN architecture for 45 epochs decreasing the NLOS rejection probability $1-\lambda$ from 1 by steps of $0.2$ every 9 epochs until reaches 0. In this manner, the first batches contain only LOS samples, whereas during the last $9$ epochs the ratio between LOS and NLOS will be the same as the one in the original unbiased empirical distribution.  We also consider the opposite strategy, namely exposing the NN to batches of hard samples first. This is obtained by the same sampling procedure but with the role of NLOS and LOS swapped. We term this alternative anti-curriculum as it starts from the hardest instances. As a natural baseline, we also consider the standard unbiased sampling procedure. For each case, we train the same model architecture proposed in Sec. \ref{subsec:Architecture} using the loss function with $\beta=0.8$. Distinct training dynamics result in different convergence time and final accuracy values.  In Fig. \ref{fig:CurrLearningCurves}, we plot the evolution of the accuracy metrics, averaged over 10 repetitions, of the standard, curriculum and anti-curriculum learning procedures. In terms of convergence time the curriculum learning strategy outperforms both the standard and anti-curriculum sampling schemes as it quickly plateaus to higher accuracy levels. The final performance, also averaged over 10 runs, is reported in Table \ref{tab:Curr}. The curriculum learning strategy improves by 2\% the top-1 throughput ratio and by 1.2\%  the top-5 throughput ratio compared to standard learning. On the other hand, anti-curriculum learning has a detrimental effect on the performance, resulting in a performance loss of 4\% and 1.7\%  in terms of top-1 and top-5 throughput ratio, respectively.

\begin{figure}[!ht]
    \centering
    \includegraphics[width=0.95\linewidth]{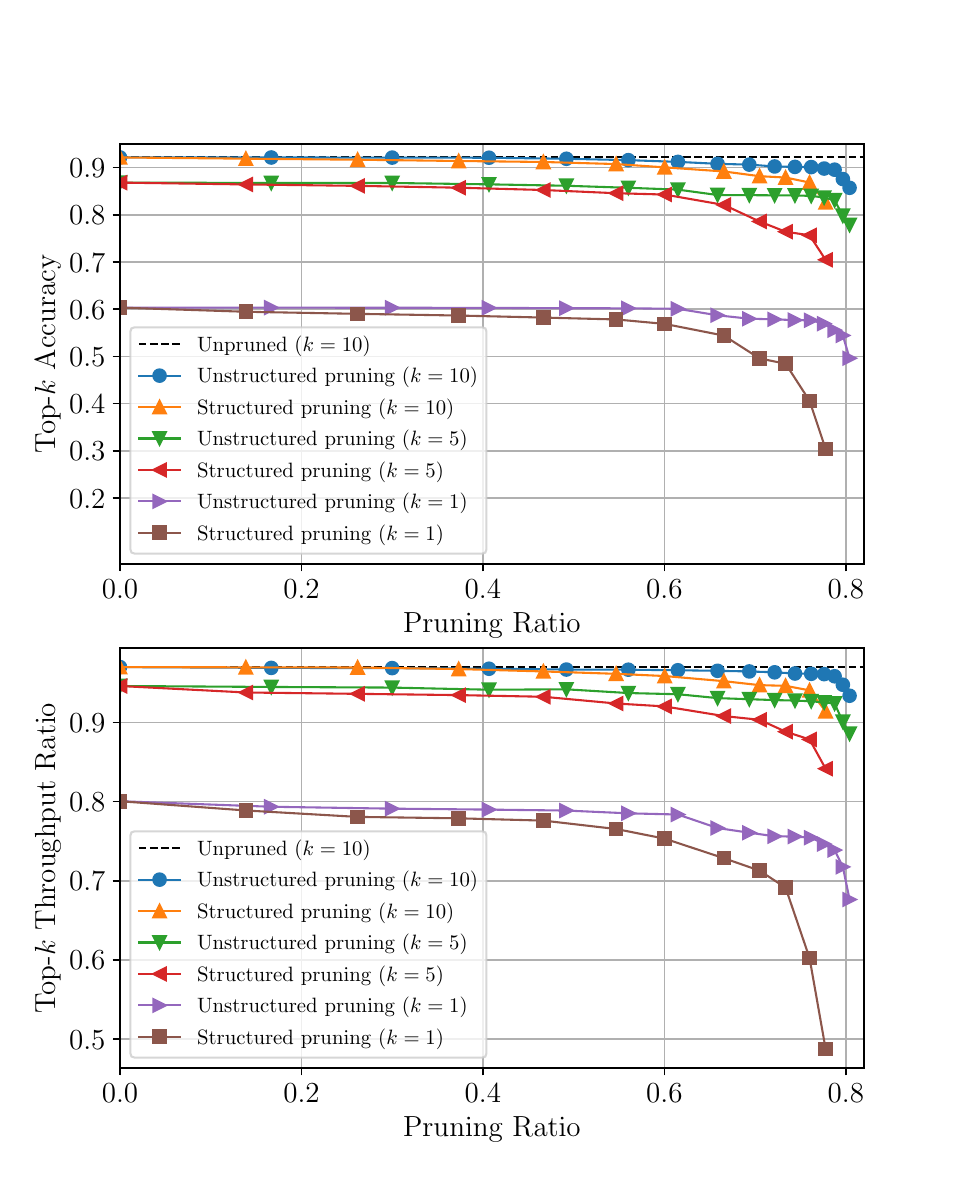}
    \caption{Accuracy and throughput ratio as a function of pruning ratio.} 
    \label{fig:pruning009}
\end{figure}

\begin{table}[t]
\centering
\caption{Performance improvement by the proposed non-local attention approach.}
\begin{tabular}{|c|c|c|c|c|} 
\hline
$\phi(\cdot)$  & $A(1)$ & $A(5)$  & $T(1)$ & $T(5)$ \\ 
\hline
\hline
Embedded & \textbf{58.3\%} & \textbf{86.7\%}  & \textbf{78.7\%} &\textbf{94.3\%} \\
\hline
Gaussian & 57.9\% & 86.5\%  & 77.7\% &  94.3\% \\
\hline
Dot &  57.7\% & 86.7\% & 77.7\% & 94.1\%  \\
\hline

Without NLA & 57.1\% & 86.0\%  & 75.9\%& 92.9\% \\
\hline

\end{tabular}
\vspace{1em}
\label{tab:NLA}
\end{table}
\subsection{Non-local Attention} In the following we consider augmenting the NN architecture of Sec. \ref{subsec:Architecture} by adding one non-local attention module located after the fifth convolutional layer of the LIDAR processing branch. This design choice represents the best trade-off between the performance gain and additional computational burden. In fact, even if multiple non-local blocks generally lead to better performance, our NN is reasonably shallow and it does not display a significant improvement when more than one non-local block is instantiated. Additionally, we did not observe any meaningful variation in the final performance by placing the non-local block at different depths in the network. On the other hand, the number of floating point operations (FLOPs) required for a forward pass through the non-local module depends on its position within the NN.  In fact, high level representations at deeper stages of the NN have lower dimension resulting in modest computational effort when processed by the non-local block. In particular, the output of the fifth convolutional layer is a tensors of size $5\times50\times5$, that is further sub-sampled using a max pooling kernel of size $1\times2\times2$. This operation preserves the non-local behaviour but it translates in  computational cost of only 250k floating point operations.

We test three popular choices for the non-local operations denoted by $\phi(\cdot)$: Gaussian, embedded Gaussian and dot product pair-wise functions.
For each, we train the model minimizing the proposed loss for $\beta=0.8$ and the standard sampling procedure. In this way we are able to assess the gain that the non-local attention module brings independently of curriculum learning. We average the final performance over 10 runs and report the average results in Table \ref{tab:NLA}. Across the three different choices of $\phi(\cdot)$ we do not observe great variability, with the non-local attention using embedded Gaussian slightly outperforming the others. Nonetheless, compared to the same model without non-local attention we measure an improvement of $2.8\%$ and $1.4\%$ in terms of top-1 and top-5 throughput ratios, respectively. 
\begin{table*}[h]
\centering
\caption{Performance comparison between the proposed NN model and previous works.}
\begin{tabular}{|c|c|c|c|c|c|c|c|c|} 
\hline
Model & $A(1)$ & $T(1)$ & $A(5)$ & $T(5)$ & $A(10)$ & $T(10)$ & FLOP count & \# params. \\ 
\hline
\hline
 \cite{8, 9} & $31.5 \pm 2.6\%$ & $46.1 \pm 2.6\%$ & $71.9 \pm 2.2\%$ & $76.1 \pm 1.9\%$ & $83.9 \pm 0.9\%$ & $86.1 \pm 0.8\%$ & $179.01 \times 10^6$ & $403677$ \\
\hline
\cite{IPC} & $52.3\pm 1.9\%$ & $70.3 \pm 2.6\%$ & $85.3 \pm 0.9\%$ & $90.8 \pm 1.5\%$ & $91.1 \pm 0.3 \%$ & $94.7 \pm 0.6\%$ & $1.72 \times 10^6$ & $7462$ \\
\hline
Proposed & $59.5 \pm 0.5\%$ & $79.9 \pm 0.8\%$ & $87.0 \pm 0.3\%$ & $94.6 \pm 0.8\%$ & $92.2 \pm  0.2 \% $ & $96.9 \pm 0.6\%$ & $4.55 \times 10^6$ & $30872$ \\
\hline
\end{tabular}
\label{tab:results}
\vspace{-0.2cm}
\end{table*}

\subsection{Pruning}

\label{subsection:pruning_exp}

We consider network pruning to further reduce the computational and storage cost of deploying the proposed NN solution. Before we prune any of the weights, we first pretrain the network, following the strategy described in Sec. \ref{subsec:Architecture}. After the pretraining is finished, we run multiple iterations of pruning by first discarding a portion of the remaining weights with the lowest $L_1$-norm and fine-tuning the network by following the same training strategy that we used for pretraining. We evaluate both the unstructured and structured pruning methodologies introduced in  Sec. \ref{subsection:pruning}. The results are reported in Fig. \ref{fig:pruning009}, where we plot the top-$k$ accuracy and top-$k$ throughput ratio of the pruned model against the pruning ratio. The pruning ratio is defined here as the fraction of the weights removed from the unpruned model. 

The unstructured pruning achieves better accuracy and throughput ratio at larger pruning rates, compared to structured pruning. This is due to the fact that unstructured pruning can be more precise in removing weights since it works by removing single weights rather than entire columns or rows of the weight matrix. Nevertheless, in unstructured pruning, weight matrices of the linear layers are only partially sparsified, which is computationally sub-optimal compared to removing entire rows or columns as structured pruning does. In both cases we find that the pruned model maintains excellent predictive performance even for large pruning ratios. In fact, it is possible to obtain a top-10 throughput ratio above $96.5\%$, despite pruning $60\%$ of the weights in the case of structured pruning, and $95.9\%$ with unstructured pruning. A graceful performance degradation also happens in terms of accuracy. Therefore, we conclude that pruning represents a viable option to further reduce the computational and storage footprint of learned models while maintaining excellent predictive capabilities.

\begin{figure}
    \centering
    \includegraphics[width=\linewidth]{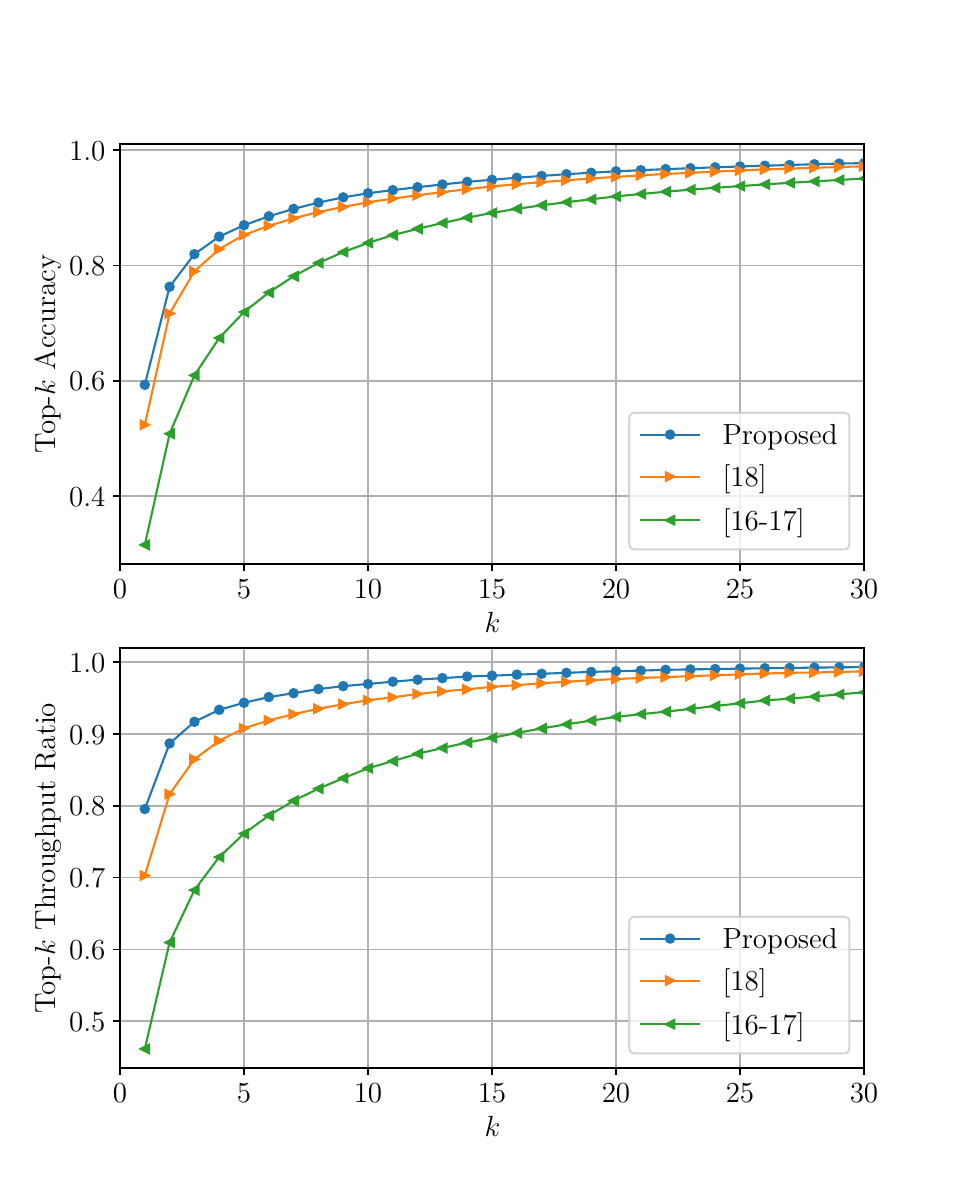}
    \caption{Accuracy and throughput ratio performance curves comparing the proposed solution and previous work.}
    \label{fig:Acc_Thr}
\end{figure}


\begin{figure*}[]
\centering
\minipage{0.33\textwidth}%
  \includegraphics[height=1.05\linewidth]{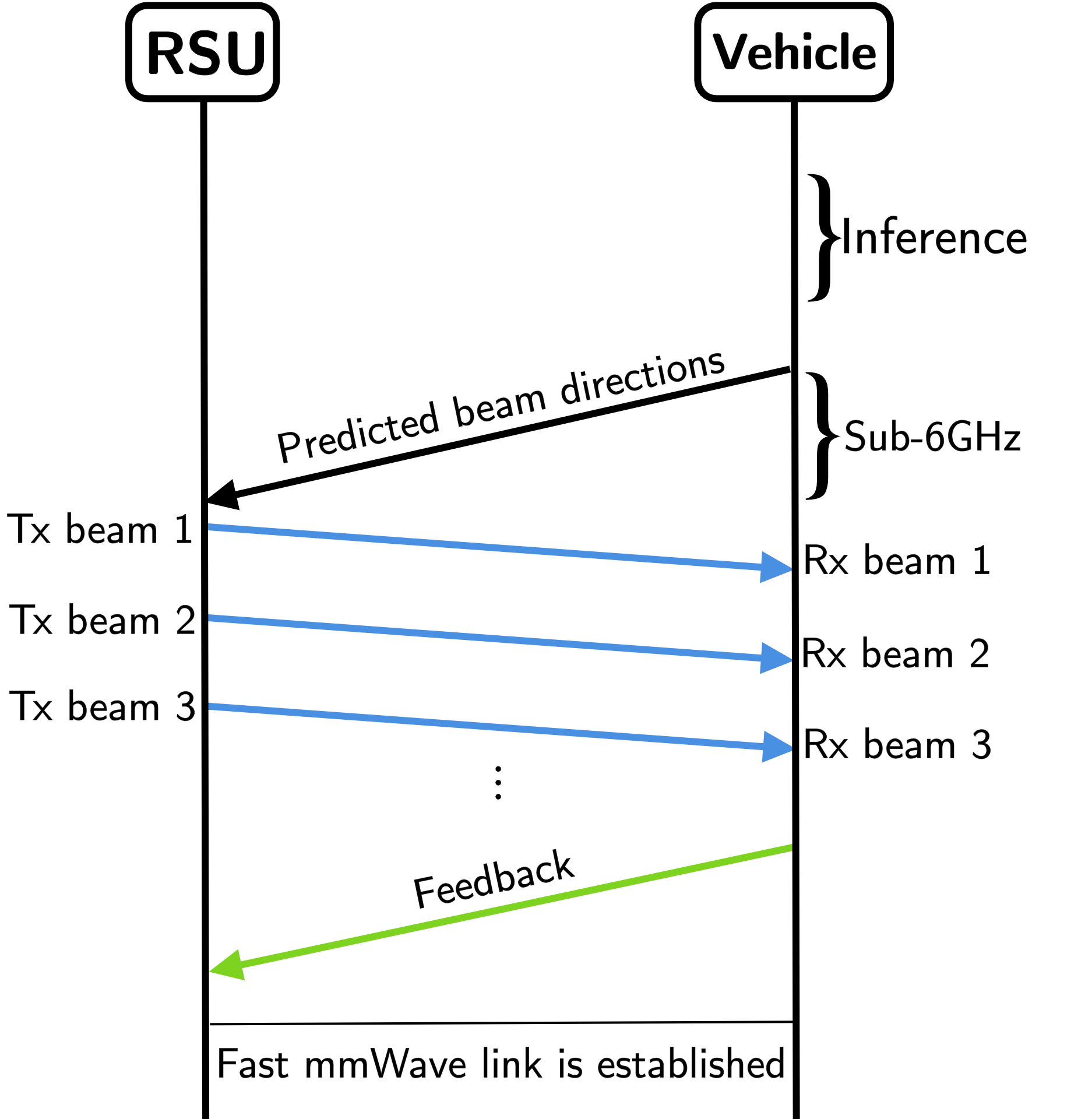}
  \subcaption{  Proposed \label{fig:Time1}}
\endminipage
\minipage{0.33\textwidth}
  \includegraphics[height=1.05\linewidth]{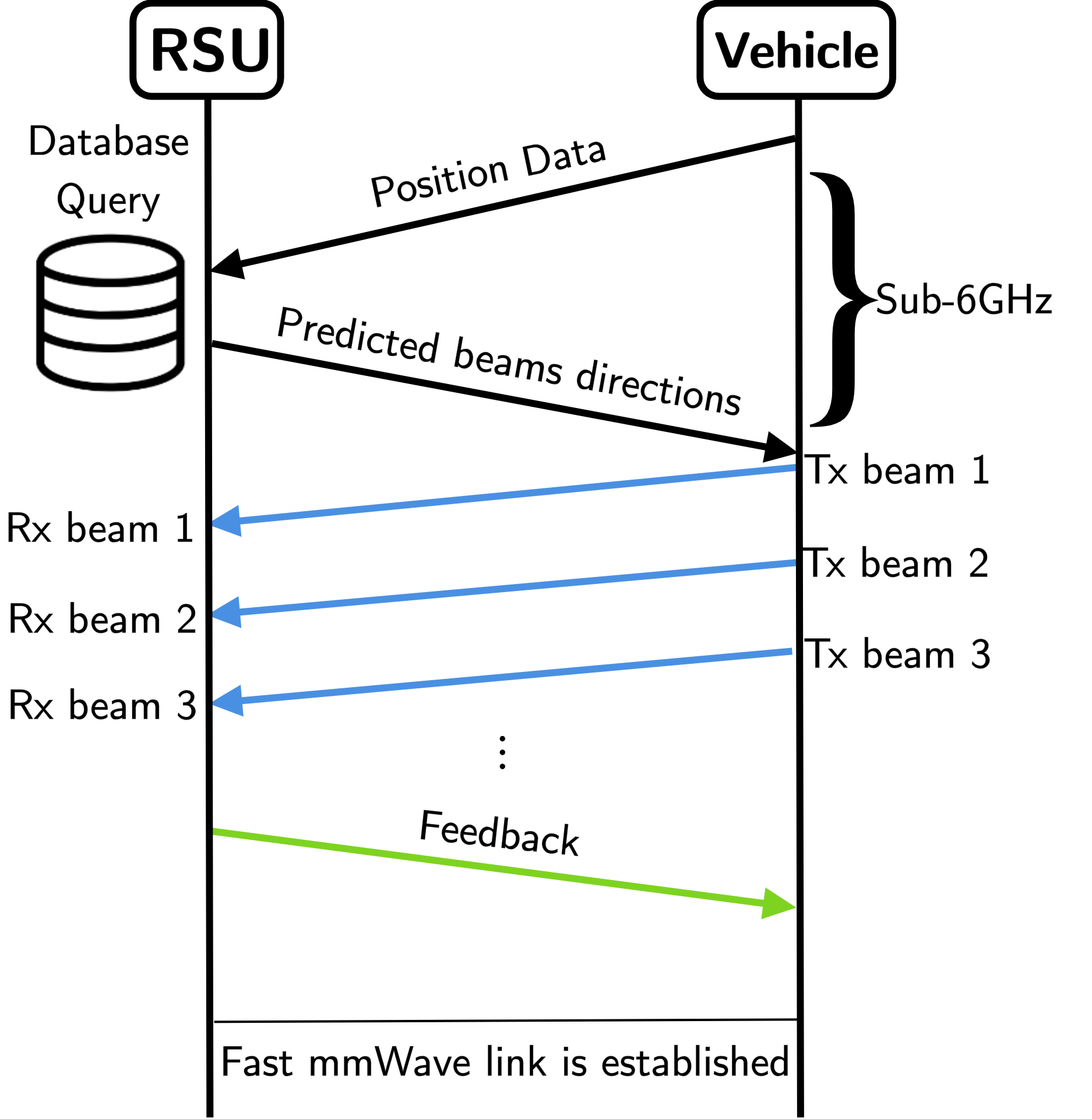}
  \subcaption{Inverse Fingerprint\label{fig:Time2}}
\endminipage
\minipage{0.33\textwidth}%
  \includegraphics[height=1.05\linewidth]{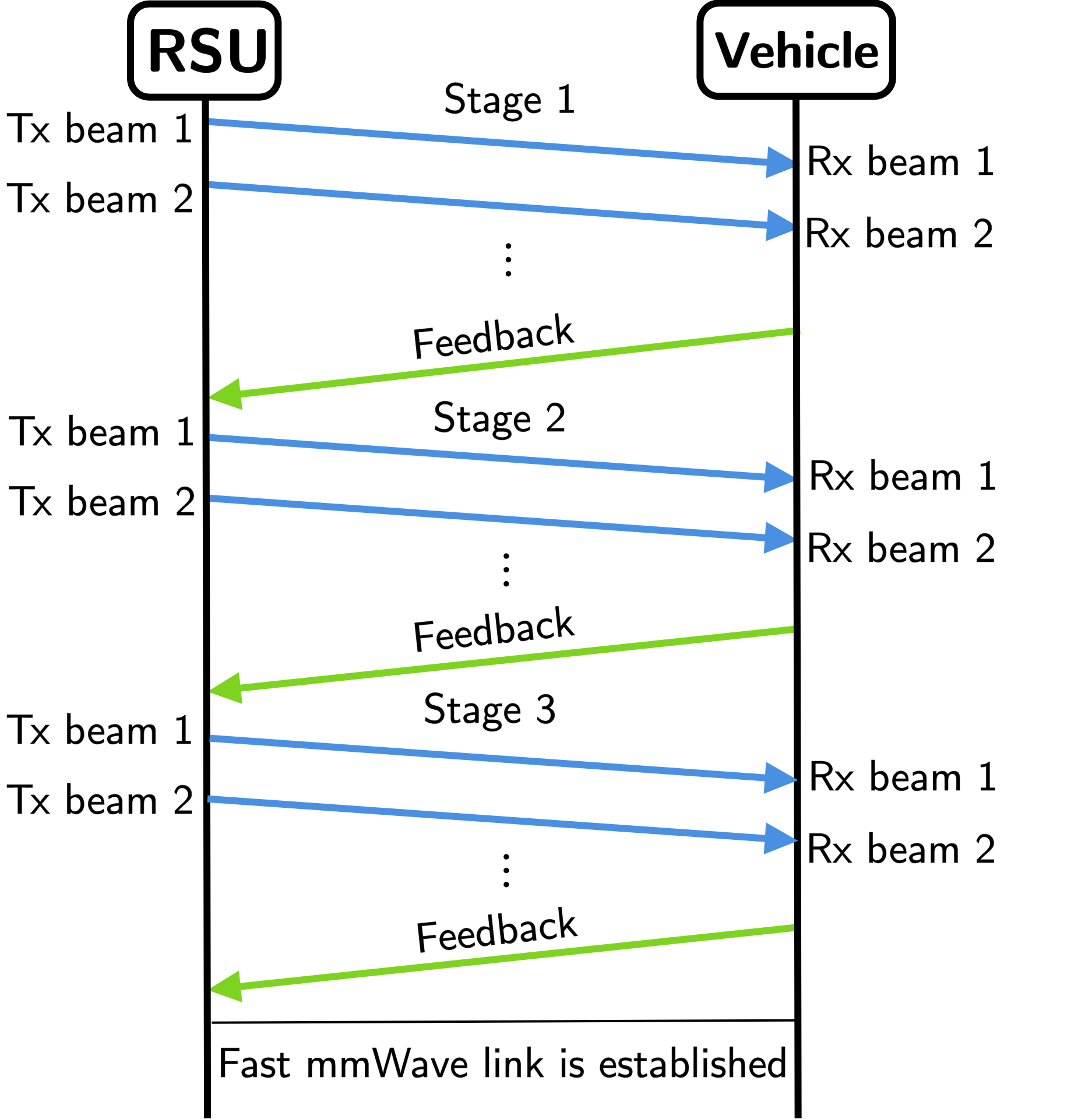}
  \subcaption{  3-Stage Hierarchical Search \label{fig:Time3}}
\endminipage

\caption{\color{black}Timing diagrams of (a) the proposed NN-based beam alignment protocol using LIDAR and position information, (b) the inverse fingerprinting scheme using position information only \cite{va2017inverse} and (c) a 3-stage hierarchical search \cite{noh2017multi}. }\label{fig:TimeDiagram}
\end{figure*}

\subsection{Comparison with LIDAR and Position-Aided Neural Beam Selection Techniques}
In the following we evaluate and compare the proposed final solution that includes all the above improvements against two state of the art data-driven approaches. The first baseline model we consider is the 13-layer NN presented in \cite{8,9} and the second one is the lighter version recently proposed in \cite{IPC} that comprises 8 layers. In Table \ref{tab:results} we compare the accuracy and throughput achieved by these models along with their number of trainable parameters and FLOPs. These results are averaged over 10 Monte Carlo rounds of training and reported with the corresponding 95\% confidence intervals. For a fair comparison, we have reported results for our unpruned architecture. The number of trainable parameters and FLOP count directly relate to the computational and memory footprint of the models, and therefore, are of interest in order to asses their deployment feasibility. Compared to the larger architecture in \cite{8,9}, our model requires only $7.6\%$ of the trainable parameters and $2.5\%$ of the floating point operations to perform the forward pass. At the same time our model is comparable to \cite{IPC} in terms of computational and storage cost; the difference can be further reduced using the pruning techniques proposed in Sec. \ref{subsection:pruning}. This renders the proposed solution also applicable in distributed training scenarios with memory and computationally constrained devices. Our model consistently outperforms both baselines in term of top-$k$ accuracy and throughput ratio metrics, which are reported in Fig. \ref{fig:Acc_Thr} for $k\in[0,30]$. Our model yields a striking $79.9\%$ top-$1$ throughput ratio, harnessing a great portion of the available rate without any search procedure. The two competing baselines attain only $46.1\%$ and $70.3\%$ top-1 throughput ratios respectively. At the same time, in order to ensure a $95\%$ expected throughput ratio our model needs to sweep on average 6 beams, greatly reducing the beam search overhead. As a comparison, the two baselines requires 28 and 11 beams, respectively. In terms of NLOS performance, our model outperfoms both alternatives providing an average top-1 throughput ratio of $79.0\%$, improving upon the two baselines by $25\%$ and $4\%$, respectively. Finally, our proposed model achieves a tighter 95\% confidence interval in comparison with \cite{8,9,IPC} specifically for smaller $k$. This is very favourable in practice as it ensures more reliable performance guarantees for different instances of training and deployment of our model. 
\color{black}
\color{black}


\color{black}
\section{NN-based Beam Alignment Protocol}\label{protocol}
In the previous section we showed that our solution outperforms state-of-the-art NN-based baselines that infer the most prominent beam directions leveraging LIDAR and position information. In this section, we present a protocol for NN-based beam alignment to be integrated into the existing 5G NR standard, through which we show the feasibility of the proposed solution in practical scenarios considering the delay and computation overhead associated with NN-based inference. We also provide comparisons with alternative beam alignment protocols.

\subsection{Integration with 5G NR}
The beam sweeping procedure in 5G NR is based on synchronization signal blocks (SSB), which are used to probe the channel when using each specific beam. In order to sweep multiple angular directions, a sequence of SSBs are time-multiplexed into a SSB burst. The length of SSB bursts is fixed to $T_{br}=5$ms and the number of SSBs that fit into a single burst depends on the operating frequency. In particular, for communication under 3GHz, the number of SSBs in a burst is $4$, while it is $8$ for communication at frequencies in the 3 to 6 GHz range, and can be up to 64 SSBs in a single burst for higher frequencies \cite{dahlman20205g}. In our scenario, the operating frequency is 60GHz, leading to a single SSB duration of $T_{SSB}=T_{br}/64=0.078$ms. 

Fig. \ref{fig:Time1} depicts our design that integrates the proposed NN-based beam alignment procedure with the existing beam search protocol of 5G NR. In the proposed design, the vehicle that wishes to communicate, predicts a subset of best mmWave beams using its pre-trained NN module fed with its readily available LIDAR and position data. Afterwards, it sends to the RSU the transmitter beam indices corresponding to the NN-predicted subset of candidate beams. These candidate beam indices are transmitted over a sub-6GHz link which is assumed to be available. Upon reception at the RSU side, the standard 5G NR SSB-based beam sweeping procedure is performed over the NN-predicted subset of candidate beams. Once the search is complete, the vehicle reports its measurements to the RSU and the mmWave link is established. The total time associated with the beam search process in this design is given by 

\begin{equation}
T_{NN} = T_{Inf} + T_{CBC} + K \times T_{SSB} + T_{FB},
\label{eqn:Prot1}
\end{equation}
where $T_{NN}$ is the total time required for NN-based beam alignment, $K$ is the number of candidate beams suggested by the NN, and $T_{Inf}$, $T_{CBC}$, and $T_{FB}$ denote the time associated with NN inference, communicating the NN-predicted candidate beams over the sub-6GHz link, and the feedback, respectively. The overhead introduced by the NN-based approach amounts to the inference time and the time required to communicate the NN-predicted candidate beams over the sub-6GHz link. Measuring the beam inference time (including data pre-processing and forward path of the NN) using a consumer grade CPU, we have $T_{Inf} \preceq 1$ms. We expect even smaller inference times in case of dedicated hardware. The $T_{CBC}$ and $T_{FB}$ terms depend on the data rate of the sub-6GHz link, but are typically very small (i.e., $T_{CBC}, T_{FB} \preceq 0.1$ms). 


\subsection{Comparison with Alternative Beam Alignment Protocols}

Herein, we compare the performance of the proposed protocol against alternative beam alignment protocols, and show that the benefits of processing LIDAR data by the NN to predict the candidate beams justifies the initial inference overhead it introduces. 

\subsubsection{\textbf{Inverse fingerprinting}}

The inverse fingerprinting beam alignment protocol proposed in \cite{va2017inverse}, and depicted in Fig. \ref{fig:Time2}, exploits the position data to query a site-specific database, which is available at the RSU, and contains channel measurements along with the vehicle position information. According to the inverse fingerprinting protocol, the vehicle sends the position information to the RSU. Upon reception, the RSU determines the candidate beam directions using the previously measured channel fingerprints from users in the same location. The RSU then sends back the candidate SSB beam sweeping signals. Once all the candidate beam directions are probed, the RSU updates the local database and feedbacks the measurements to the vehicle in order to establish a fast mmWave link. As a result, the total time $T_{IF}$ required for the fingerprinting is given by

\begin{equation}
T_{IF} = T_{GPS} + T_{Qr} + T_{CBC} + K \times T_{SSB} + T_{FB},
\label{eqn:Prot2}
\end{equation}
where $K$ is the number of candidate beams returned by querying the database, and $T_{GPS}$, $T_{Qr}$, $T_{CBC}$, and $T_{FB}$ denote the time associated with communicating the GPS coordinates, querying the database at the RSU, communicating the resulting candidate beams, and the feedback, respectively. In Eq. (\ref{eqn:Prot2}), the $T_{GPS}$ and $T_{CBC}$ terms depend on the data rate of the sub-6GHz link and $T_{Qr}$ depends on the processing power available for querying the database. 

\subsubsection{\textbf{Hierarchical beam selection}}
An alternative class of algorithms carry out coordinated search without making use of any side information. The simplest scheme in this class is the coordinated exhaustive search that sequentially tests all possible beams in a predefined order. A more efficient alternative is the coordinated hierarchical beam search which leverages multi-resolution codebooks at the receiver and transmitter side to sweep beams with an increasing angular resolution as depicted in Fig. \ref{fig:Time3}. Hierarchical beam search greatly reduces the search space but it is known to be prone to compounding errors that can mislead the search.

\begin{figure}[]
    \centering
    \includegraphics[width=1\linewidth]{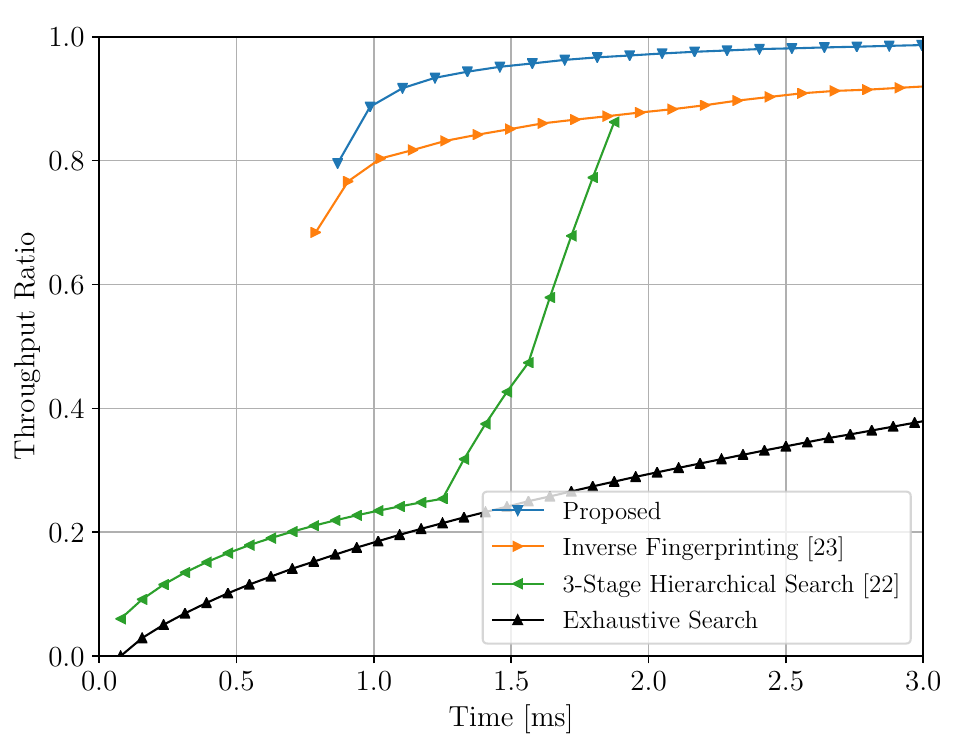}
    \caption{\color{black}{Throughput ratio versus beam sweeping time. The proposed and the inverse fingerprint schemes curves are obtained evaluating the schemes for different values of $K$. They introduce a modest overhead to the overall beam training but they are able to perform a more efficient informed search compared to the hierarchical and exhaustive sweeping algorithms.}} 
    \label{fig:Time}
\end{figure}

In Fig. \ref{fig:Time} we compare the above beam alignment schemes and report the fraction of throughput obtained as the sweeping progresses. In our comparisons, we consider all the components affecting the search time of each protocol as provided in Eqs. (\ref{eqn:Prot1}) and (\ref{eqn:Prot2}). In particular, we benchmark the algorithms over the s009 portion of data while we use the s008 dataset to train the neural network model and to build the site-specific database for inverse fingerprints. We assume that a sub-6GHz link is available with a data rate of 0.1 Mb/sec and that each coordinate is quantized into 16 bits. We also endow the RSU and the vehicle with consumer-grade computing capabilities and accordingly set $T_{inf}=0.75$ ms and $T_{Qr}=0.3$ ms. 

In Fig. \ref{fig:Time}, we increase the number of candidate beams $K$ to achieve different points on the throughput ratio versus time curves, for our proposed approach and inverse fingerprinting. For hierarchical beam search, we consider the RSU and vehicle equipped with a DFT-based three-level codebook as in \cite{noh2017multi}. We denote the codebooks at the RSU by $\{\mathcal{C}^1_t,\mathcal{C}^2_t,\mathcal{C}^3_t\}$ and at the vehicle by $\{\mathcal{C}^1_r,\mathcal{C}^2_r,\mathcal{C}^3_r\}$. We fix the corresponding codebook sizes to be, respectively, $\{8,16,32\}$ at the RSU and $\{2,4,8\}$ at the vehicle. The RSU and vehicle first sweep the beams in $\mathcal{C}^1_t\times\mathcal{C}^1_r$ and then refine the search over the angular sector that results in the largest channel gain using the subset of beams in $\mathcal{C}^2_t\times\mathcal{C}^2_r$. The refinement procedure is then repeated using the beams in $\mathcal{C}^3_t\times\mathcal{C}^3_r$ to produce the final beam. For this specific choice of codebooks the search procedure ends after sweeping through 24 beam pairs. As expected, the throughput ratio sharply improves when switching from a lower resolution to a higher resolution one.

Overall, the delay introduced by the data-driven methods is very modest, and, at the same time, utilizing sensor information yields high throughput beam pairs with minimal search. For this reason, despite the initial offset due to the processing of the side information and the transmission of the predicted best beam directions, the inverse fingerprinting and the proposed method outperform both the exhaustive and hierarchical search methods. In this scenario, our proposed approach almost halves the search time compared to the inverse fingerprinting scheme to attain a 90$\%$ throughput ratio. This shows the benefits of using the additional LIDAR side information in comparison with the inverse fingerprinting scheme that only uses the position data.

\color{black}


\section{Conclusions}\label{sec:Conclusions}
We have proposed a supervised learning scheme for efficient mmWave beam selection that exploits side information in the form of {\color{black}{LIDAR and position}} data to reduce the beam search overhead. Our approach significantly outperformed the state of the art in terms of beam classification accuracy and resulting throughput. We have introduced a non-local attention block to improve the performance, specifically for the more challenging NLOS scenarios. Additionally, we have proposed a curriculum learning strategy and a novel loss function inspired by knowledge distillation, which improved the training speed and the accuracy of the final solution. Our NN-based beam selection scheme was able to harness almost 79.9\% of the available throughput without any beam search, just by utilizing the LIDAR and location side information. {\color{black}In a typical mmWave V2I scenario, our proposed method considerably reduces the beam search time required to achieve a desired throughput, in comparison with the inverse fingerprinting and hierarchical beam selection schemes.} Our NN architecture can be further simplified by pruning up to 60\% of its parameters without considerable performance loss, hence, rendering it suitable for practical deployment. 


\bibliographystyle{IEEEtran}
\bibliography{refs}

\begin{thebibliography}{10}
\providecommand{\url}[1]{#1}
\csname url@samestyle\endcsname
\providecommand{\newblock}{\relax}
\providecommand{\bibinfo}[2]{#2}
\providecommand{\BIBentrySTDinterwordspacing}{\spaceskip=0pt\relax}
\providecommand{\BIBentryALTinterwordstretchfactor}{4}
\providecommand{\BIBentryALTinterwordspacing}{\spaceskip=\fontdimen2\font plus
\BIBentryALTinterwordstretchfactor\fontdimen3\font minus
  \fontdimen4\font\relax}
\providecommand{\BIBforeignlanguage}[2]{{%
\expandafter\ifx\csname l@#1\endcsname\relax
\typeout{** WARNING: IEEEtran.bst: No hyphenation pattern has been}%
\typeout{** loaded for the language `#1'. Using the pattern for}%
\typeout{** the default language instead.}%
\else
\language=\csname l@#1\endcsname
\fi
#2}}
\providecommand{\BIBdecl}{\relax}
\BIBdecl

\bibitem{choi2016millimeter}
J.~Choi, V.~Va, N.~Gonzalez-Prelcic, R.~Daniels, C.~R. Bhat, and R.~W. Heath,
  ``Millimeter-wave vehicular communication to support massive automotive
  sensing,'' \emph{IEEE Communications Magazine}, vol.~54, no.~12, pp.
  160--167, 2016.

\bibitem{va2016impact}
V.~Va, J.~Choi, and R.~W. Heath, ``The impact of beamwidth on temporal channel
  variation in vehicular channels and its implications,'' \emph{IEEE
  Transactions on Vehicular Technology}, vol.~66, no.~6, pp. 5014--5029, 2016.

\bibitem{wei2018initial}
L.~Wei, Q.~Li, and G.~Wu, ``Initial access techniques for {5G} {NR}: Omni/beam
  {SYNC} and {RACH} designs,'' in \emph{2018 International Conference on
  Computing, Networking and Communications (ICNC)}.\hskip 1em plus 0.5em minus
  0.4em\relax IEEE, 2018, pp. 249--253.

\bibitem{hur2011multilevel}
S.~Hur, T.~Kim, D.~J. Love, J.~V. Krogmeier, T.~A. Thomas, and A.~Ghosh,
  ``Multilevel millimeter wave beamforming for wireless backhaul,'' in
  \emph{2011 IEEE Globecom Workshops}.\hskip 1em plus 0.5em minus 0.4em\relax
  IEEE, 2011, pp. 253--257.

\bibitem{1}
V.~{Va}, J.~{Choi}, T.~{Shimizu}, G.~{Bansal}, and R.~W. {Heath}, ``Inverse
  multipath fingerprinting for millimeter wave {V2I} beam alignment,''
  \emph{IEEE Transactions on Vehicular Technology}, vol.~67, no.~5, pp.
  4042--4058, 2018.

\bibitem{brambilla2019inertial}
M.~Brambilla, M.~Nicoli, S.~Savaresi, and U.~Spagnolini, ``Inertial sensor
  aided mm{W}ave beam tracking to support cooperative autonomous driving,'' in
  \emph{2019 IEEE International Conference on Communications Workshops (ICC
  Workshops)}.\hskip 1em plus 0.5em minus 0.4em\relax IEEE, 2019, pp. 1--6.

\bibitem{8277251}
I.~{Mavromatis}, A.~{Tassi}, R.~J. {Piechocki}, and A.~{Nix}, ``Beam alignment
  for millimeter wave links with motion prediction of autonomous vehicles,'' in
  \emph{Antennas, Propagation RF Technology for Transport and Autonomous
  Platforms 2017}, 2017, pp. 1--8.

\bibitem{7}
N.~{González-Prelcic}, R.~{Méndez-Rial}, and R.~W. {Heath}, ``Radar aided
  beam alignment in mm{W}ave {V2I} communications supporting antenna
  diversity,'' in \emph{2016 Information Theory and Applications Workshop
  (ITA)}, 2016, pp. 1--7.

\bibitem{5}
T.~{Nitsche}, A.~B. {Flores}, E.~W. {Knightly}, and J.~{Widmer}, ``Steering
  with eyes closed: mm-{W}ave beam steering without in-band measurement,'' in
  \emph{2015 IEEE Conference on Computer Communications (INFOCOM)}, 2015, pp.
  2416--2424.

\bibitem{6}
A.~{Ali}, N.~{González-Prelcic}, and R.~W. {Heath}, ``Millimeter wave
  beam-selection using out-of-band spatial information,'' \emph{IEEE
  Transactions on Wireless Communications}, vol.~17, no.~2, pp. 1038--1052,
  2018.

\bibitem{Sub6G1}
M.~{Alrabeiah} and A.~{Alkhateeb}, ``Deep learning for mm{W}ave beam and
  blockage prediction using sub-6 {GHz} channels,'' \emph{IEEE Transactions on
  Communications}, vol.~68, no.~9, pp. 5504--5518, 2020.

\bibitem{Sub6G2}
I.~Chafaa, R.~Negrel, E.~V. Belmega, and M.~Debbah, ``Federated channel-beam
  mapping: from sub-6ghz to mmwave,'' in \emph{IEEE WCNC 2021 Workshop:
  Distributed Machine Learning}, 2021.

\bibitem{vision}
V.~M. {De Pinho}, M.~L.~R. {De Campos}, L.~U. {Garcia}, and D.~{Popescu},
  ``Vision-aided radio: User identity match in radio and video domains using
  machine learning,'' \emph{IEEE Access}, vol.~8, pp. 209\,619--209\,629, 2020.

\bibitem{alrabeiah2020millimeter}
M.~Alrabeiah, A.~Hredzak, and A.~Alkhateeb, ``Millimeter wave base stations
  with cameras: Vision-aided beam and blockage prediction,'' in \emph{2020 IEEE
  91st Vehicular Technology Conference (VTC2020-Spring)}.\hskip 1em plus 0.5em
  minus 0.4em\relax IEEE, 2020, pp. 1--5.

\bibitem{Alkhateeb}
W.~{Xu}, F.~{Gao}, S.~{Jin}, and A.~{Alkhateeb}, ``{3D} scene-based beam
  selection for mm{W}ave communications,'' \emph{IEEE Wireless Communications
  Letters}, vol.~9, no.~11, pp. 1850--1854, 2020.

\bibitem{8}
A.~{Klautau}, N.~{González-Prelcic}, and R.~W. {Heath}, ``{LIDAR} data for
  deep learning-based mm{W}ave beam-selection,'' \emph{IEEE Wireless
  Communications Letters}, vol.~8, no.~3, pp. 909--912, 2019.

\bibitem{9}
M.~{Dias}, A.~{Klautau}, N.~{González-Prelcic}, and R.~W. {Heath}, ``Position
  and {LIDAR}-aided mm{W}ave beam selection using deep learning,'' in
  \emph{2019 IEEE 20th International Workshop on Signal Processing Advances in
  Wireless Communications (SPAWC)}, 2019, pp. 1--5.

\bibitem{IPC}
M.~B. Mashhadi, M.~Jankowski, T.-Y. Tung, S.~Kobus, and D.~Gündüz,
  ``Federated mm{W}ave beam selection utilizing {LIDAR} data,'' \emph{IEEE
  Wireless Communications Letters}, vol.~10, no.~10, pp. 2269--2273, 2021.

\bibitem{challenge}
\emph{AI/ML in 5G}, accessed April 2021,
  \url{https://www.itu.int/en/ITU-T/AI/challenge/2020/Pages/default.aspx}.

\bibitem{ranking}
\emph{AI/ML in 5G}, accessed April 2021,
  \url{https://www.itu.int/en/ITU-T/AI/challenge/2020/Pages/results.aspx}.

\bibitem{hinton2015distilling}
G.~Hinton, O.~Vinyals, and J.~Dean, ``Distilling the knowledge in a neural
  network,'' \emph{NIPS 2014 Deep Learning Workshop.}, 2014.

\bibitem{noh2017multi}
S.~Noh, M.~D. Zoltowski, and D.~J. Love, ``Multi-resolution codebook and
  adaptive beamforming sequence design for millimeter wave beam alignment,''
  \emph{IEEE Transactions on Wireless Communications}, vol.~16, no.~9, pp.
  5689--5701, 2017.

\bibitem{va2017inverse}
V.~Va, J.~Choi, T.~Shimizu, G.~Bansal, and R.~W. Heath, ``Inverse multipath
  fingerprinting for millimeter wave {V2I} beam alignment,'' \emph{IEEE
  Transactions on Vehicular Technology}, vol.~67, no.~5, pp. 4042--4058, 2017.

\bibitem{raymobtime}
\emph{Raymobtime}, accessed November 2020,
  \url{https://www.lasse.ufpa.br/raymobtime/}.

\bibitem{raymobtime1}
A.~{Klautau}, P.~{Batista}, N.~{González-Prelcic}, Y.~{Wang}, and R.~W.
  {Heath}, ``{5G MIMO} data for machine learning: application to beam-selection
  using deep learning,'' in \emph{2018 Information Theory and Applications
  Workshop (ITA)}, 2018, pp. 1--9.

\bibitem{KD2}
P.~I. Yonglong~Tian, Dilip~Krishnan, ``Contrastive representation
  distillation,'' in \emph{Proceedings of the International Conference on
  Learning Representations, {ICLR} 2020, April 2020, {Addis Ababa},
  {Ethiopia}}.

\bibitem{bengio2009curriculum}
Y.~Bengio, J.~Louradour, R.~Collobert, and J.~Weston, ``Curriculum learning,''
  in \emph{Proceedings of the 26th annual international conference on machine
  learning}, 2009, pp. 41--48.

\bibitem{Curriculum2}
G.~Hacohen and D.~Weinshall, ``On the power of curriculum learning in training
  deep networks,'' in \emph{Proceedings of the 36th International Conference on
  Machine Learning, {ICML} 2019, 9-15 June 2019, Long Beach, California,
  {USA}}.

\bibitem{attention1}
A.~{Buades}, B.~{Coll}, and J.~. {Morel}, ``A non-local algorithm for image
  denoising,'' in \emph{2005 IEEE Computer Society Conference on Computer
  Vision and Pattern Recognition (CVPR'05)}, vol.~2, 2005, pp. 60--65 vol. 2.

\bibitem{attention2}
X.~{Wang}, R.~{Girshick}, A.~{Gupta}, and K.~{He}, ``Non-local neural
  networks,'' in \emph{2018 IEEE/CVF Conference on Computer Vision and Pattern
  Recognition}, 2018, pp. 7794--7803.

\bibitem{pruning}
Y.~LeCun, J.~S. Denker, and S.~A. Solla, ``Optimal brain damage,'' in
  \emph{Advances in Neural Information Processing Systems}, 1990, pp. 598--605.

\bibitem{pruning2}
S.~Han, J.~Pool, J.~Tran, and W.~Dally, ``Learning both weights and connections
  for efficient neural network,'' in \emph{Advances in Neural Information
  Processing Systems}, vol.~28, 2015.

\bibitem{dahlman20205g}
E.~Dahlman, S.~Parkvall, and J.~Skold, \emph{5G NR: The next generation
  wireless access technology}.\hskip 1em plus 0.5em minus 0.4em\relax Academic
  Press, 2020.

\end{thebibliography}

\begin{IEEEbiography}[{\includegraphics[width=1in,height=1.25in,clip,keepaspectratio]{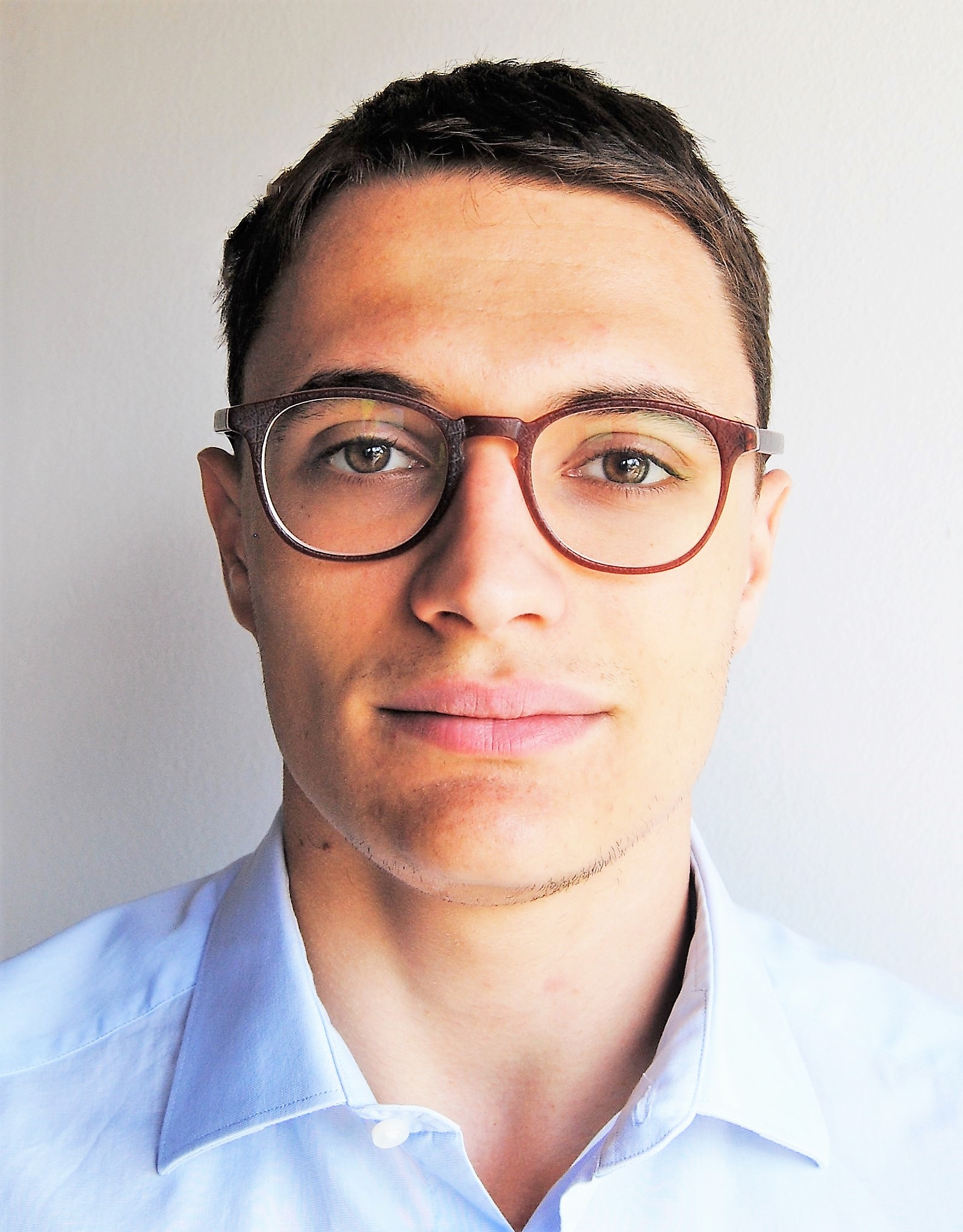}}]%
{Zecchin Matteo}
(Student Member, IEEE) received the BSc degree in Information Engineering from Università di Padova, Padova, Italy, in 2017 and the MSc degree in Telecommunications Engineering from Università di Padova and National Taiwan University in 2019, both cum laude. He is currently an Early Stage Research in the H2020 Project Windmill and a Ph.D student within the Department of Communication Systems, EURECOM, Sophia Antipolis, France.
\end{IEEEbiography}

\begin{IEEEbiography}[{\includegraphics[width=1in,height=1.25in,clip,keepaspectratio]{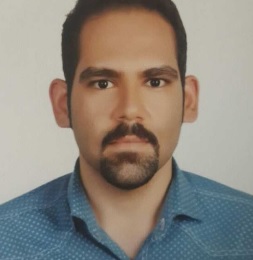}}]%
{Mahdi Boloursaz Mashhadi}
[S’14-M’20] 
received a dual major B.S. degree (Hons.) in electrical and industrial engineering and the M.S. and Ph.D. degrees in electrical engineering from Sharif University of Technology (SUT), Tehran, Iran, in 2011, 2013 and 2018, respectively. He is currently a lecturer (assistant professor) at the 5GIC \& 6GIC, Institute for Communication Systems (ICS), University of Surrey, UK. He previously worked as a post-doctoral research associate at the Intelligent Systems and Networks (ISN) research group, Imperial College London, UK, and as a visiting research associate at the University of Central Florida (UCF), and Queens’ University, Canada. His research interests include the areas of wireless communications, machine learning, and sparse/statistical signal processing. His awards include the Best Paper Award at the EWDTS 2012 conference, and several national, regional, and IEEE grants. He serves as a reviewer for IEEE Transactions on Wireless Communications, IEEE Transactions on Communications, and IEEE Transactions on Signal Processing.
\end{IEEEbiography}

\begin{IEEEbiography}[{\includegraphics[width=1in,height=1.25in,clip,keepaspectratio]{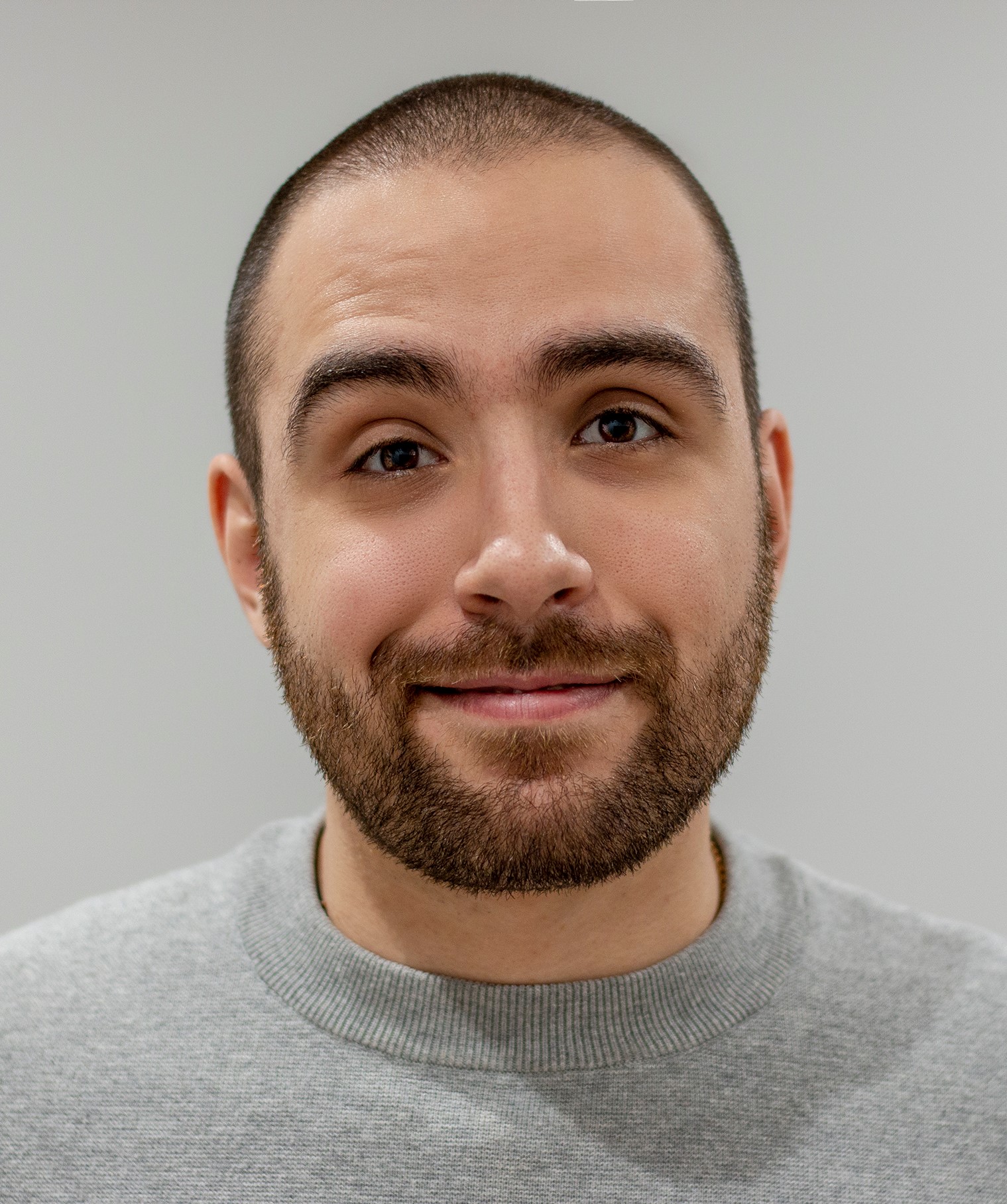}}]%
{Mikolaj Jankowski}
(Graduate Student Member, IEEE) received the B.Sc. degree in control and robotics from the Warsaw University of Technology in 2016, and the M.Sc. degree in control systems from Imperial College London in 2018. He is currently pursuing the Ph.D. degree with Intelligent Systems and Networks Group, under the supervision of Prof. Gündüz and Prof. Mikolajczyk. His main research areas are machine learning, computer vision, image processing and information theory. He is currently working on data compression optimized for various machine learning tasks and applying machine learning to improve wireless communications.
\end{IEEEbiography}

\begin{IEEEbiography}[{\includegraphics[width=1in,height=1.25in,clip,keepaspectratio]{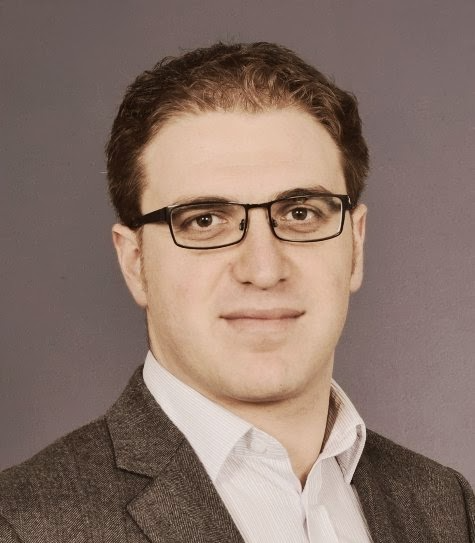}}]%
{Deniz G\"und\"uz}
[S’03-M’08-SM’13] 
received the B.S. degree in electrical and electronics engineering from METU, Turkey in 2002, and the M.S. and Ph.D. degrees in electrical engineering from NYU Tandon School of Engineering (formerly Polytechnic University) in 2004 and 2007, respectively. After his PhD, he served as a postdoctoral research associate at Princeton University, as a consulting assistant professor at Stanford University, and as a research associate at CTTC in Barcelona, Spain. ln Sep. 2012, he joined the Electrical and Electronic Engineering Department of Imperial College London, UK, where he is currently a Professor of Information Processing, and serves as the deputy head of the Intelligent Systems and Networks Group. He is also a part-time faculty member at the University of Modena and Reggio Emilia, Italy, and has held visiting positions at University of Padova (2018-2020) and Princeton University (2009-2012).

His research interests lie in the areas of communications and information theory, machine learning, and privacy. Dr. Gündüz is a Distinguished Lecturer for the IEEE Information Theory Society (2020-22). He is an Area Editor for the IEEE Transactions on Communications and the IEEE Journal on Selected Areas in Communications (JSAC). He also serves as an Editor of the IEEE Transactions on Wireless Communications. He is an organiser and co-chair of the 2021 London Symposium on Information Theory. He is the recipient of the IEEE Communications Society - Communication Theory Technical Committee (CTTC) Early Achievement Award in 2017, a Starting Grant of the European Research Council (ERC) in 2016, and several best paper awards. 
\end{IEEEbiography}

\begin{IEEEbiography}[{\includegraphics[width=1in,height=1.25in,clip,keepaspectratio]{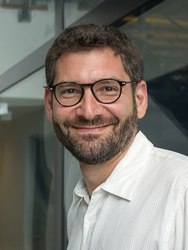}}]%
{Marios Kountouris}
[S'04--M'08--SM'15]
received the diploma degree in electrical and computer engineering from NTUA, Athens, Greece in 2002 and the M. S. and Ph.D. degrees in electrical engineering from T\'{e}l\'{e}com Paris, France in 2004 and 2008, respectively. He is currently a Professor and Chair PI on Advanced Wireless Systems at EURECOM, France. Prior to his current appointment, he has held positions at Huawei Paris Research Center, France, CentraleSup\'{e}lec, France, The University of Texas at Austin, USA, and Yonsei University, S. Korea. He was awarded a consolidator grant by the European Research Council (ERC) in 2020 on goal-oriented semantic communication. He has served as Associate Editor for the IEEE Transactions on Wireless Communications, the IEEE Transactions on Signal Processing, and the IEEE Wireless Communication Letters. He has received several awards, including the 2020 IEEE ComSoc Young Author Best Paper Award, the 2016 IEEE ComSoc CTTC Early Achievement Award, the 2013 IEEE ComSoc Outstanding Young Researcher Award for the EMEA Region, the 2012 IEEE SPS Signal Processing Magazine Award, the IEEE SPAWC 2013 Best Paper Award, and the IEEE Globecom 2009 CT Best Paper Award.
\end{IEEEbiography}

\begin{IEEEbiography}[{\includegraphics[width=1in,height=1.25in,clip,keepaspectratio]{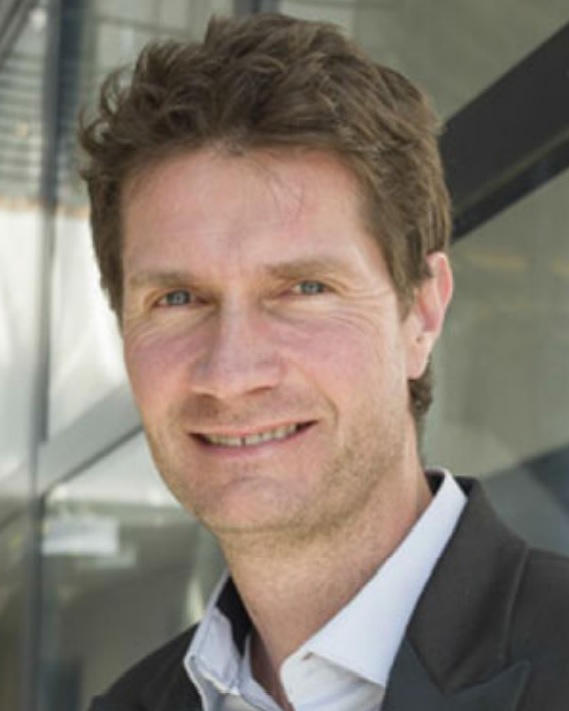}}]%
{David Gesbert}
(Fellow, IEEE) received the Ph.D. degree from the Ecole Nationale Superieure des Telecommunications, France, in 1997. He is currently a Professor and the Head of the Communication Systems Department, EURECOM. From 1997 to 1999, he was with the Information Systems Laboratory, Stanford University. He was the Founding Engineer of Iospan Wireless Inc., a Stanford spin off pioneering MIMO-OFDM (currently Intel). Before joining EURECOM in 2004, he was with the Department of Informatics, University of Oslo, as an Adjunct Professor. He has published about 340 articles and 25 patents, some of them winning the 2019 ICC Best Paper Award, the 2015 IEEE Best Tutorial Paper Award (Communications Society), the 2012 SPS Signal Processing Magazine Best Paper Award, the 2004 IEEE Best Tutorial Paper Award (Communications Society), the 2005 Young Author Best Paper Award for Signal Processing Society journals, and the paper awards at conferences 2011 IEEE SPAWC and 2004 ACM MSWiM. He has been the Technical Program Co-Chair for ICC2017. He was named a Thomson-Reuters Highly Cited Researchers in computer science. He was on the Advisory Board of the Huawei European Research Institute. He is a Board Member for the OpenAirInterface (OAI) Software Alliance. In 2015, he was awarded the ERC Advanced Grant “PERFUME” on the topic of smart device communications in future wireless networks. Since early 2019, he has been heading the Huawei-funded Chair on Advanced Wireless Systems Towards 6G Networks. In 2020, he was awarded funding by the French Interdisciplinary Institute on Artificial Intelligence for a Chair in the area of AI for the future IoT.
\end{IEEEbiography}

\end{document}